\documentclass[manuscript=true,screen,review=false,nonacm]{acmart}
\AtBeginDocument{%
  }

\usepackage{adjustbox}
\usepackage{graphicx}
\usepackage{listings,xcolor}
\usepackage{subcaption}
\usepackage{tabularx}
\usepackage{longtable}
\usepackage{placeins}
\usepackage[many]{tcolorbox}
\usepackage{wrapfig}

\definecolor{codebg}{RGB}{248,248,248}
\lstdefinestyle{llmmini}{
  language=Python,
  basicstyle=\ttfamily\footnotesize,
  keywordstyle=\color{blue!60!black}\bfseries,
  commentstyle=\color{green!45!black},
  stringstyle=\color{purple!70!black},
  frame=single, frameround=tttt, rulecolor=\color{black!25},
  backgroundcolor=\color{codebg},
  showstringspaces=false, columns=fullflexible, keepspaces=true,
  breaklines=true, tabsize=2,
  numbers=left, numberstyle=\color{black!45}\scriptsize,
  numbersep=6pt, stepnumber=1, firstnumber=1,
  xleftmargin=1.4em, framexleftmargin=1.4em,
  aboveskip=4pt, belowskip=0pt
}

\newcommand{\findingsbox}[1]{
\begin{tcolorbox}[breakable,width=\linewidth,
boxrule=0pt, leftrule = 6pt, top=1pt, bottom=1pt, left=1pt,right=1pt, 
colback=gray!20,colframe=gray!60]
\textbf{Insight:} #1
\end{tcolorbox}
}

\iftrue           % to not display the ideas use iffalse
 \newcommand{\Idea}[1]{\textcolor{purple}{[#1]}}
 \else
 \newcommand{\Idea}[1]{}
 \fi

\iffalse       % to put back in original text color use iffalse
 \newcommand{\Update}[1]{\textcolor{blue}{#1}}
 \else
 \newcommand{\Update}[1]{#1}
 \fi

\newcommand{\ie}{\textit{i.e., }}
\newcommand{\eg}{\textit{e.g., }}
\newcommand{\etal}{\textit{et al. }}

\makeatletter
\newcommand{\labeltext}[2]{%
  \@bsphack
  \csname phantomsection\endcsname % in case hyperref is used
  \def\@currentlabel{#1}{\label{#2}}%
  \@esphack
}

   
\acmSubmissionID{}
\setcopyright{none}               % removing copyright info
\renewcommand\footnotetextcopyrightpermission[1]{}  % to remove footer text
\begin{document}

%%
%% The "title" command has an optional parameter,
%% allowing the author to define a "short title" to be used in page headers.
% \title{Sustainable Serving of Large Language Models: Energy-Performance Trade-offs in Multi-Request Workflow-Based Inference}
\title{Characterizing Performance--Energy Trade-offs of Large Language Models in Multi-Request Workflows }

% \title{Towards Sustainable LLM Deployment: A Measurement Study of Multi-Request Workflow-Level Energy Efficiency}

% \author{Anonymous Authors}
\author{Md. Monzurul Amin Ifath}
% \authornote{Both authors contributed equally to this research.}
\email{monzurul.amin@dal.ca}
\orcid{0009-0002-1373-273X}
\affiliation{%
  \institution{Dalhousie University}
  \city{Halifax}
  \state{NS}
  \country{Canada}
}
\author{Israat Haque}
% \authornote{Both authors contributed equally to this research.}
\email{israat@dal.ca}
\orcid{0000-0003-4450-3358}
\affiliation{%
  \institution{Dalhousie University}
  \city{Halifax}
  \state{NS}
  \country{Canada}
}

%%
%% By default, the full list of authors will be used in the page
%% headers. Often, this list is too long, and will overlap
%% other information printed in the page headers. This command allows
%% the author to define a more concise list
%% of authors' names for this purpose.
% \renewcommand{\shortauthors}{Monzurul et al.}

\begin{abstract}
Large language models (LLMs) are increasingly deployed in applications forming multi-request workflows like document summarization, search-based copilots, and multi-agent programming. While these workflows unlock richer functionality, they also amplify latency and energy demand during inferences. Existing measurement and benchmarking efforts either focus on assessing performance of LLM inference systems or consider single-request evaluations, overlooking workflow dependencies and cross-request interactions unique to multi-request workflows. Moreover, the energy usage of such interdependent LLM calls is not explored in-depth.
%Moreover, existing investigations provide limited insight into the energy usage of such interdependent LLM calls.
%

% In this paper, we characterize performance--energy trade-offs in multi-request LLM inference. 
To address these gaps, this paper presents the first systematic characterization of performance--energy trade-offs in multi-request LLM inference. We develop and evaluate four representative workloads that capture sequential, interactive, agentic, and composite patterns common in modern deployments. Using an empirical NVIDIA A100 testbed with state-of-the-art serving systems (vLLM and Parrot), we systematically analyze how key energy knobs (e.g., input-output length, batch size, and GPU power cap) reshape latency, throughput, and component-level (e.g., CPU, GPU, and DRAM) energy use. Our findings reveal that batch size is the most impactful lever, though its benefits are highly workload dependent. While optimal batching benefits workloads with large shared prompts, it is ineffective for sequential summarization and only partially effective for mutli-agent coding. GPU power capping provides modest but predictable savings, while output length induces linear energy scaling with limited efficiency gains. We further demonstrate that engine-level optimizations in vLLM (e.g., continuous batching, PagedAttention) maintain higher GPU utilization and efficiency, especially for decode-heavy workloads, while Parrot’s workflow-aware scheduling achieves lower energy consumption under stringent power constraints. These findings offer actionable guidelines for developers and system operators in designing performance- and energy-aware LLM serving systems in emerging multi-request workflows.
% These findings can be attributed as guidelines for developers and system operators in realizing performance-energy-aware LLM systems in emerging multi-request workflows. 
\end{abstract}

% \ccsdesc[300]{Hardware~Impact on the environment}
% \ccsdesc[300]{Information systems~Language models}
% \ccsdesc[300]{Social and professional topics~Sustainability}


% \setcopyright{cc}
% \setcctype{by}
% \acmJournal{POMACS}
% \acmYear{2026} \acmVolume{10} \acmNumber{1} \acmArticle{7} \acmMonth{3} \acmPrice{}\acmDOI{10.1145/3788089}

%%
%% This command processes the author and affiliation and title
%% information and builds the first part of the formatted document.
\maketitle

\section{Introduction}
\label{intro}

% \textcolor{red}{workflow: the importance LLMs in modern AI applications -> these applications can be based on single or multi-request tasks -> introduce multi-request tasks -> talk how their usage creates skyrocket compute and energy need, especially in inference -> how a group of works completely missed considering energy -> how another group partially investigated the energy aspects -> this paper fills that important gap -> what we did? while doing what challenges we encountered and solved (it will be our methodology) -> evaluations answer those research qsns not method, imo -> which workload, and why? which serving systems, and why? -> what's our key findings -? summarize the contribution and observations --> organization of the paper }

Large language models (LLMs) have rapidly become the backbone of modern AI applications, powering productivity assistants, conversational copilots, knowledge-intensive search, and autonomous multi-agent frameworks~\cite{brown2024large,chen2024more,wu2023autogen,zhuge2024language}, which are widely adopted in consumer-facing chatbots, domain-specific copilots in programming, healthcare, and education. The key enabler of these applications is LLM \emph{inference}, the process of generating outputs from a pre-trained model given an input prompt. The inference can be single- or multi-request depending on the chain of invocation, such as translating a short text segment (\textit{single-request}) vs. summarizing a long document by invoking a pipeline of segment-by-segment summarization (\textit{multi-request}). The adoption of multi-request workflows is becoming a norm in the LLM application domain, \eg conversational copilots that iteratively refine search results under safety and context constraints~\cite{microsoftCopilot,googleGemini}, and agentic programming frameworks (\eg MetaGPT~\cite{metaGPT}, AutoGen~\cite{autogen}) that coordinate specialized roles across architect, coder, and reviewer agents.  

Unlike LLM model training, which is computationally intensive but episodic, inference of these multi-request workflows are continuously invoked in respective applications, making inference the dominant factor in resource consumption (\eg computation and energy) of LLM systems. 
A plethora of systems \cite{agrawal2024taming,alizadeh2024llm,kwon2023efficient,li2023alpaserve,patel2024splitwise,yu2022orca} have focused on optimizing computational demand of LLMs without accounting on their energy demand and consequences. However, AI inference accounts for 60-70\% of AI power use in hyperscale infrastructures~\cite{wu2022sustainable}, with a single chatbot interaction consuming up to $10\times$ more energy than a typical web search~\cite{iea2024efficiency}. 
The energy consumption of LLM inference is influenced by a set of well-known knobs, including workload-level parameters such as batch size and input-output length, model-level parallelism strategies (\eg data, pipeline, or tensor parallelism), and system-level controls such as GPU power capping~\cite{fernandez-etal-2025-energy,maliakel2025investigating}. 

Several attempts have been made to characterize the performance (throughput and latency) of single-request inference systems or to explore performance–energy trade-offs over a subset of the available control knobs. In the case of multi-request workflows, some studies focus only on the performance of LLM serving systems, the middleware that connects high-level LLM applications to low-level inference backends. However, multi-request workflows exhibit unique characteristics such as dependencies among LLM calls, heterogeneous prompt structures, and varying complexity levels along critical paths, which reduces the scope of batching, prefix reuse, and latency reduction, respectively. Thus, the performance--energy characterization must consider the interplay between the application layer, the serving system, and the backend (\ie LLM engine) while incorporating the control knobs at the workload, model, and system level. Thus, comprehensive characterization will enable the design of sustainable LLM systems, which is missing in existing measurements.   

We conduct the first comprehensive study of performance--energy trade-offs for mutli-request workflow-based LLM inferences. To do so, we consider four representative workloads of document chain summarization, LLM-powered search, multi-agent coding, and composite (see examples in Figure~\ref{fig:workflows}). We specifically choose these four workloads because they capture the dominant multi-request patterns observed in practice (\eg LangChain~\cite{langchain2025}): sequential pipelines (document summarization), interactive copilots (LLM-powered search), agent-driven directed acyclic graphs (DAGs) (multi-agent coding), and heterogeneous mixes of workloads (composite). Across these workloads, we systematically evaluate three key energy knobs (input-output length, batch size, and GPU power capping) because they have been consistently identified as the most critical levers for shaping the performance--energy trade-off in LLM inference. Batch size directly influences GPU utilization and throughput efficiency, input-output length determines both prefill and decode costs, and GPU power capping provides a practical system-level mechanism to bound energy usage under latency or other service-level objectives (SLOs).

Conducting this study required addressing several practical challenges. Workflow orchestration introduced strong dependencies that complicated cache management and concurrency control. Ensuring accurate and reproducible energy measurement demanded fine-grained monitoring of hardware components (CPU, GPU, and DRAM) while isolating background interference. Serving system heterogeneity further introduced implementation and deployment variations, requiring careful tuning and validation. To mitigate these challenges, we explicitly modeled dependencies within workflow definitions and validated execution order, fixed concurrency levels to maintain workload consistency, and standardized prompt templates across runs. Each serving system was containerized to ensure isolation. Under stable thermal conditions, we utilized Zeus~\cite{zeusGit} to measure component-level energy consumption, for repeated trials to ensure dependability. All workloads were deployed and evaluated on a controlled NVIDIA A100 testbed for consistency and reproducibility.

The evaluation first investigates which knobs are most influential on workload efficiency, how the most impactful knob shapes performance--energy trade-offs across applications, and how different serving system designs behave under the most energy-intensive workload. Our findings highlight that \emph{batch size} is the most impactful knob, reduces energy per token by up to $2.6\times$ for LLM-based search workloads, but negatively impacts sequential summarization when dependencies dominate. The output length exhibits near-linear energy scaling with limited efficiency gains, while GPU power capping offers modest but predictable energy savings at the expense of throughput. 

Finally, we compare two widely adopted serving systems, vLLM~\cite{kwon2023efficient} and Parrot~\cite{lin2024parrot}, under the most energy-intensive workload (document chain summarization). We focus on vLLM and Parrot as they embody two contrasting design philosophies. vLLM embodies high-throughput, engine-level optimizations, while Parrot adopts workflow-aware scheduling through semantic variables that capture inter-request dependencies. Our results show that vLLM’s low-level optimizations sustain higher GPU utilization, leading to consistently lower energy consumption. In contrast, Parrot achieves relative gains only under strict power constraints, leveraging its workflow-aware scheduling to balance CPU-GPU activity. For example, in the document chain summarization with batch size 16, vLLM reduced GPU stall time and achieved consistently better utilization, offering up to 28\% lower energy consumption than Parrot. We summarize our main contributions as follows.

\begin{itemize}
    \item We develop and evaluate four multi-request workloads under a controlled NVIDIA A100-based testbed with fine-grained CPU, GPU, and DRAM energy monitoring. 
    
    \item We conduct the first comprehensive analysis of performance--energy trade-offs in workflow-based LLM inference, focusing on three critical energy knobs (input-output length, batch size, and GPU power capping). 
    
    \item We compare two serving systems, vLLM and Parrot, to demonstrate how engine-level optimizations contrast with workflow-aware scheduling in shaping energy efficiency, particularly under high-load.
    
    \item Our findings highlight that (i) batch size is the most impactful but workload-dependent knob, reducing energy per token by up to $2.6\times$ in search while degrading sequential summarization; (ii) output length induces proportional energy scaling with limited efficiency gains; and (iii) GPU power capping provides modest, yet predictable, energy savings.
\end{itemize}

The remainder of this paper is structured as follows. Section~\ref{sec:background} introduces the foundations of LLM serving systems and energy control knobs. Section~\ref{sec:related-work} surveys related work. Section~\ref{sec:deployment_exp_setup} describes our experimental setup and measurement methodology. Section~\ref{sec:evaluation} presents empirical results addressing three research questions. We discuss our findings further in Section~\ref{sec:discussion}. We make concluding remarks in Section~\ref{sec:conclusion}.

\section{Background}
\label{sec:background}

\begin{wrapfigure}{r}{0.4\linewidth}
  \centering
  % \vspace{-10pt} % adjust vertical placement if needed
  \includegraphics[width=\linewidth,height=0.25\textheight]{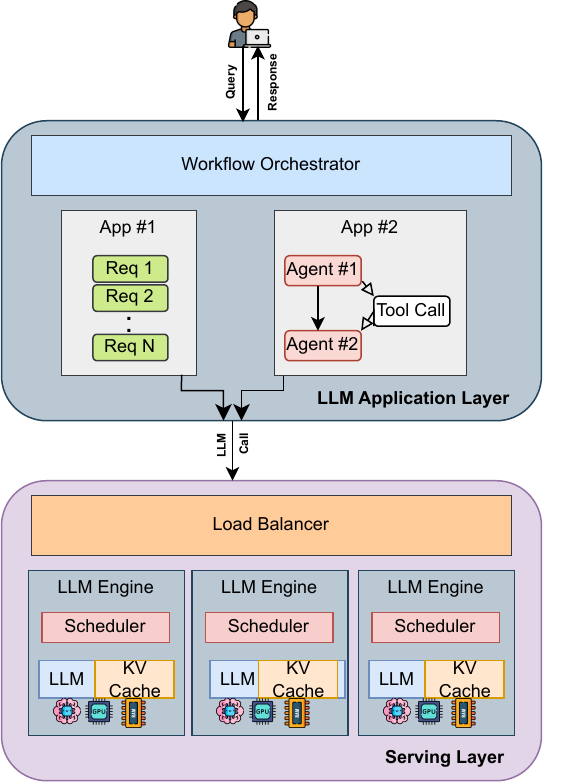}
  \caption{Architectural layers of LLM inference, spanning the LLM application layer (top) and the LLM serving layer (bottom).}
  \label{fig:llm_workflow_layer}
  \vspace{-5pt}
\end{wrapfigure}

This section starts with the architectural components of LLM inference (see Figure~\ref{fig:llm_workflow_layer}), where we first describe the LLM application layer, which captures workflow-level properties and orchestration of multi-request applications. Next, we examine the LLM serving layer, which connects applications to low-level inference backends and manages system-level optimizations. We conclude with an overview of energy measurement and control knobs that frame our study of performance--energy trade-offs in workflow-based LLM serving.

\noindent{\textbf{LLM Inference:}}
Modern LLMs primarily adopt the transformer architecture~\cite{vaswani2017attention}, which comes in three main variants: encoder-only, decoder-only, and encoder-decoder \cite{minaee2024large}. For generative tasks, decoder-only architectures (\eg GPT~\cite{brown2020language}, LLaMA~\cite{touvron2023llama}) are the most common, which produce output tokens sequentially by conditioning input tokens and contextual representations embedded in the model parameters. This sequential token generation (aka \textit{auto-regressive}) processes one token at a time.

LLM inference typically consists of two distinct stages: \textit{prefill} and \textit{generate} (\textit{decode}). In the prefill stage, the input prompt is tokenized and the model computes the attention relationships among these tokens. In the generate stage, the model predicts the next token iteratively, each time leveraging the previously generated tokens. To avoid recomputing attention for past tokens~\cite{yu2022orca}, the intermediate key-value representations are stored in a dedicated memory structure called \textit{KV cache}. The prefill stage exhibits high parallelizability, lead to computation bottleneck (\ie being compute-bound), while the auto-regressive decoding stage is primarily limited by the memory-bandwidth due to repeated access to model weights and the KV cache \cite{patel2024splitwise}. 
\noindent{\textbf{LLM Application Layer:}} 
LLM-based applications often extend beyond a single model invocation, forming
multi-step workflows where multiple LLM calls are connected through well-defined control and data dependencies ~\cite{luo2025autellix}. For example, long document summarization uses chain style pipelines, LLM-based search iteratively reformulates and synthesizes answers, and multi-turn planning-execution loops in multi-agent\footnote{An LLM agent refers to an autonomous module driven by an LLM that performs a specific role (\eg planning, coding, or reviewing) and communicates through structured prompts and responses within a coordinated workflow \cite{wu2023autogen}.} programming paradigms. In such workflows, the output of one stage often becomes the input to another and the overall application latency is determined by the critical path (the longest chain of dependent LLM calls determines the end-to-end completion of a workflow) across these stages. Capturing and exposing these dependencies allows LLMs to make informed scheduling decisions that prioritize tasks that unlock downstream progress ~\cite{lin2024parrot}. Orchestration frameworks (\eg LangChain~\cite{langchain2025}, LlamaIndex~\cite{llamaIndex}) provide higher-level abstraction to compose these workflows. They expose APIs for chaining model calls, integrating external tools, and coordinating multi-agent interactions, thereby generating the structured execution graphs. 
% (see examples in Figure~\ref{fig:workflows}). 
These frameworks form the core of the LLM \textit{Application Layer} and act as the glue between applications and serving systems, ensuring that workflow semantics are preserved while delegating low-level execution to serving systems in the later layer.
% \begin{figure}[!hbtp]
%  \centering
%  \includegraphics[width=0.6\linewidth,height=0.4\textheight,keepaspectratio]{pictures/LLM layer block dig.pdf}
%  \caption{Architectural Layers of LLM Inference.}
%  \label{fig:llm_workflow_layer}
% \end{figure}

\noindent{\textbf{LLM Serving Layer:}}
In modern LLM-based deployments, the serving layer functions as the critical middleware connecting high-level applications (\eg agents, multi-request workflows) to low-level inference backends (\eg GPU runtimes, quantized engines). A well-designed serving system must take care of scheduling and multiplexing requests across constrained hardware, mediating latency-throughput trade-offs, and incorporating orchestration decisions from the application layer \cite{li2024llm,agrawal2024taming,sun2024llumnix}. In practice, serving systems are responsible for batching, queuing, resource partitioning, model routing, KV cache reusing, and in advanced cases, application-aware scheduling or DAG-level optimization. Example includes vLLM \cite{vllm}, Parrot\cite{parrot}, llama.cpp \cite{llama-cpp}, Ollama \cite{ollama}, Huggingface Text Generation Inference (TGI) \cite{hf-tgi}, SGLang \cite{sglang}.
% A summary of existing systems is shown in Table~\ref{tbl:llm_serving_comparison}, positioning their abstractions, focus, and limitations.
% \textcolor{red}{seems extra --- A central tension in \textcolor{red}{serving system design} is balancing high token throughput (\ie efficient GPU utilization) against stringent latency requirements. Systems that emphasize throughput rely on aggressive batching, which risks inflating tail latency; others prioritize low latency at the cost of utilization. Contemporary open-source and research serving systems illustrate different choices along this spectrum, particularly in how much semantic information they expose to the scheduler.}

vLLM is a high-performance inference engine and serving system focused on memory efficiency and throughput. Its key innovation, \emph{PagedAttention} \cite{kwon2023efficient}, treats the KV cache as a paged memory space, dynamically allocating and evicting pages to reduce fragmentation under heterogeneous request lengths. Combined with continuous batching (\ie dynamically admits new requests into in-progress batches) and memory reuse, vLLM achieves high throughput while maintaining competitive latency. In contrast, Parrot introduces the \emph{Semantic Variable} abstraction to expose application-level structure to the serving system. By annotating inputs and outputs as semantic variables, applications define a DAG across dependent calls. This information enables Parrot to co-schedule dependent requests, eliminate redundant prompt prefixes, and optimize directly for end-to-end latency rather than individual request times.

SGLang combines a structured generation domain specific language (DSL) with a high-performance runtime, leveraging techniques such as RadixAttention for prefix reuse, speculative decoding, and lightweight scheduling, targeting a balance between throughput and workflow flexibility. Ollama, built on top of llama.cpp, provides a user-friendly local server with quantization and model management, suitable for personal or small-scale deployments but limited in concurrency and batching. 
%The standalone llama.cpp server mode offers a minimal HTTP interface with wide portability, though it lacks advanced scheduling capabilities. 
Finally, Hugging Face TGI provides a production-ready serving layer with support for continuous batching, multi-GPU scaling, and optimized kernels, but does not incorporate fine-grained application-level semantics into its scheduling decisions. A summary of these systems is shown in Table~\ref{tbl:llm_serving_comparison}, positioning their abstractions, focus, and limitations.

\begin{table}[!hbtp]
  \centering
  \small
  \begin{tabular}{p{1.5cm} p{3cm} p{3.8cm} p{4.3cm}}
    \toprule
    \textbf{System} & \textbf{API / Abstraction} 
    % & \textbf{Key Optimizations} 
    & \textbf{Focus (Latency vs Throughput)} & \textbf{Limitations} \\
    \midrule
    vLLM & OpenAI-style text completion API 
    % & PagedAttention, continuous batching, memory reuse 
    & High throughput with modest latency & Performance in multi-call workflows is under-explored \\
    Parrot & Semantic Variables/ DAG API 
    % & Co-schedule dependent calls, prefix sharing, DAG-aware batching 
    & End-to-end latency in multi-step workflows & Requires application annotations; not tuned for pure throughput \\
    SGLang & Structured generation DSL + runtime 
    % & RadixAttention, speculative decoding, prefix reuse 
    & Balanced workflow-aware performance & Newer system; steeper programming model \\
    Ollama & Local server wrapping over llama.cpp 
    % & Quantization, lightweight serving, model management 
    & Low-latency, small-scale use & Limited batching and scale-out \\
    llama.cpp & Minimal HTTP/ OpenAI-style API 
    % & Quantization, portability, lightweight binary 
    & Simple, edge deployment & Minimal scheduling; low throughput under load \\
    HuggingFace TGI & REST/ gRPC text-gen API 
    % & Continuous batching, multi-GPU, optimized kernels 
    & Scalable throughput with reasonable latency & Heavyweight system; no workflow awareness \\
    \bottomrule
  \end{tabular}
  \caption{Comparison of representative LLM serving systems.}
  \label{tbl:llm_serving_comparison}
\end{table}
Given this landscape, our study focuses on vLLM and Parrot. vLLM serves as the benchmark for engine-level throughput and memory optimizations, establishing the baseline efficiency of single-request inference. Parrot, by contrast, represents the application-aware paradigm, where exposing cross-call dependencies enables more holistic optimization of workflow performance. Together, these systems allow us to examine both the benefits of inference-layer optimization and the additional gains unlocked by workflow-aware scheduling.

\noindent\textbf{Energy Measurement:} 
The influencing factors in the energy consumption of LLM inference includes spanning model-, workload-, system-, and platform-level categories. Model-level knobs include model size, architecture, parallelism, and optimization techniques~\cite{samsi2023words,chitty2024llm,stojkovic2025dynamollm,argerich2024measuring}, while workload-level knobs cover batch size and input-output length~\cite{you2023zeus,stojkovic2024towards,wilkins2024offline}. System-level controls include the number of GPU instances, and GPU power capping~\cite{you2023zeus,patel2024characterizing}, while platform-level factors encompass the choice of deep learning inference frameworks (\eg PyTorch, TensorFlow)~\cite{georgiou2022green}.
Among these, three knobs at the workload and system levels have been consistently highlighted as critical for controlling the performance--energy trade-off \cite{stojkovic2024towards,samsi2023words,wilkins2024offline}. 
%\textcolor{purple}{A range of factors influence the energy consumption of LLM inference, spanning model-, workload-, system-, and platform-level categories. }
First, the LLM request input-output length directly impacts computational cost and memory access, thereby shaping energy usage~\cite{wilkins2024offline,maliakel2025investigating}. 
Second, batching plays a decisive role in GPU utilization and throughput efficiency, with larger batch sizes improving amortization of compute overheads but increasing latency sensitivity~\cite{stojkovic2024towards,samsi2023words}. 
Third, GPU power capping provides a practical mechanism to bound energy consumption under varying SLOs~\cite{kakolyris2024slo,argerich2024measuring,samsi2023words}. 

%%%While these studies provide valuable insights, they largely restrict their analysis to \emph{single-request inference} scenarios. In contrast, our work focuses on multi-request, workflow-based applications and systematically evaluates the same three knobs, input-output length, batch size, and GPU power capping, in order to uncover their implications for sustainable LLM serving.

%However, this body of work has primarily examined \emph{single-request inference}, leaving the compounded effects in multi-request workflow-based applications largely unexplored. Motivated by this gap, we center our evaluation on these three knobs as the primary axes of control for sustainable LLM serving in workflow-driven scenarios.

\textit{Input-output Length:} The input length in LLM inference refers the number of tokens present in a submitted request. On the contrary, the output length is the number of tokens generated by the model in response to that request. Increased input length necessitates greater GPU parallelism, leading to extended prefill latency; conversely, longer output sequences induce more auto-regressive iterations, which consequently increase the decode phase. Typically, the output length is explicitly capped by the application, and the inference engine halts generation upon reaching this predefined token limit. This design allows precise control over the response size and latency, which is particularly important for performance-sensitive workloads.

\textit{Batch Size:} Batch size is another workload-level parameter in LLM serving refers to the number of concurrent requests grouped and processed together in a single forward pass of the model. Batching improves GPU utilization and inference throughput by reducing memory access and computation costs across multiple requests. However, this comes at the cost of increased per-request latency due to queuing and batching delays. The optimal batch size depends on the application's tolerance to latency and the nature of the workload. 
% In multi-stage workflows like map-reduce document summarization, selectively applying larger batch sizes to throughput-sensitive stages (\eg map) while minimizing batching for latency-sensitive stages (\eg reduce) can yield improved end-to-end performance.

\textit{GPU Power Capping:} Power capping provides a system controllable hardware-level parameter to control the energy footprint of LLM inference. Modern GPUs are typically configured to operate at their peak power to maximize throughput and minimize latency. However, for use cases that are not latency-critical (\eg offline summarization or scheduled analytics tasks), operating at full power is unnecessary and may lead to energy inefficiency. Power capping enforces an upper bound on GPU power draw, enabling energy savings with potentially acceptable trade-offs in performance. Intervention to enforce GPU power or frequency limits introduces measurable, non-negligible system overhead \cite{stojkovic2025dynamollm}.
% In our evaluation, we systematically vary the GPU power cap (150W to 250W, the latter being the A100's peak) and assess its impact on latency, throughput, and energy efficiency. While prior work has primarily considered individual inference requests under power capping, our study extends the scope to multi-request scenarios and diverse LLM application workflows.

% \Idea{Energy Measurement tools in data center network: software, hardware meters; add table}\\

\section{Related Work} 
\label{sec:related-work}

Related work in this section is grouped into two categories: (i) serving systems and orchestration frameworks that enable efficient execution of multi-request LLM applications and (ii) empirical studies and measurement methodologies that quantify inference energy consumption in practical deployments.\\
% along with optimization strategies that balance performance and sustainability in practical deployments.

% \Idea{LLM Inference Application}
% \Idea{Serving system related papers -> Performance focus -> Optimizations}
% \begin{itemize}
%     \item parrot, Autellix, vLLM paper, kairos
% \end{itemize}
% \Idea{Energy Knobs in ML, DL, LLM}
% \begin{itemize}
%     \item \Idea{Training -> inference -> single request -> multiple request -> agentic}
%     \item \Idea{papers:Tu(r)ning AI Green}
% \end{itemize}
% \Idea{Performance-energy tradeoff in LLM workflow scenario}
% \begin{itemize}
%     \item \Idea{papers: DynamoLLM}
% \end{itemize}

% vLLM~\cite{kwon2023efficient} introduces \textit{PagedAttention}, a mechanism that enables batching of variable-length LLM requests by allocating attention key–value states in non-adjacent memory regions. This design improves memory utilization and reduces fragmentation, allowing the system to more efficiently serve heterogeneous workloads. Parrot~\cite{lin2024parrot}, in its turn, improves end-to-end performance of LLM-based applications by leveraging semantic variables. It deploys a First-Come-First-Served (FCFS) scheduling and uses Round-Robin dispatching to distribute requests across engines. Parrot migrates KV caches across multiple LLM engines to meet request SLOs and improve overall latencies.
\noindent{\textbf{LLM Orchestrator Frameworks and Serving Systems:}} 
Modern orchestration frameworks provide developers with primitives to manage application’s control flow while deciding when to trigger multi-request workflows, invoke tools, or ask for human input. To handle complex inter-agent communication and tool execution patterns within LLM-based applications, frameworks such as LangChain \cite{langchain2025}, LllamaIndex \cite{llamaIndex}, and AutoGen \cite{autogen} have emerged. These frameworks enable developers to compose multi-step workflows by chaining LLM calls, integrating external tools (e.g., databases, APIs), and managing agent interactions through declarative or graph-based specifications \cite{bai2024digirl,schick2023toolformer,zhou2024llm}. Kairos \cite{chen2025kairos} further targets multi-agent orchestration under high concurrency, using workflow- and memory-aware scheduling. These frameworks primarily focus on responsiveness and scalability while efficiently utilizing underlying system resources. They achieve this by exposing high-level abstractions for client-side workflow specification, which simplify application composition and interaction logic. Some frameworks also consider fairness, \ie, ensuring balanced scheduling across agents or users~\cite{chaudhry2025murakkab}.

% These frameworks mostly focus on responsiveness and scalability while efficiently utilizing underlying system resources. Also, some frameworks consider fairness \ie ensuring balanced scheduling across agents or users. 
% These frameworks primarily expose high-level abstractions for client-side workflow specification, simplifying application composition and interaction logic. 
%While These frameworks improve responsiveness and scalability under load, their optimization objectives mostly remain centered on latency and resource fairness (\ie ensuring balanced scheduling across agents or users), energy-performance trade-offs remain outside their design scope.
% 

These orchestration frameworks interact with the serving systems that connect these frameworks to LLM engines. 
Recent studies have systematically benchmarked serving systems, exposing key performance trade-offs. LLM-Inference-Bench~\cite{chitty2024llm} is a benchmarking suite for various accelerators and engines such as vLLM, llama.cpp, and TensorRT-LLM, revealing heterogeneous scaling patterns across prompt lengths, model sizes, and batching. DeepSpeed-FastGen \cite{holmes2024deepspeed} improves the performance of vLLM while studies in \cite{agrawal2025evaluating,yu2022orca,kwon2023efficient} focus on other notable serving systems like HuggingFace TGI, Ollama, and SGLang. Agrawal \etal~\cite{agrawal2025evaluating} propose a serving system with fine-grained batching mechanisms and better performance.

% Researchers have explored different fine-grained serving optimizations to enhance efficiency. Yu \etal \cite{yu2022orca} develop iteration-level continuous batching while PagedAttention \cite{kwon2023efficient} improves KV cache utilization under heterogeneous request lengths. 
However, these serving systems are single-request-centric and optimize throughput and memory efficiency without exploiting cross-request relationships in multi-step workflows. Parrot \cite{lin2024parrot} fills that gap by introducing semantic variables that expose dependency graphs for co-scheduling and cache reuse, directly optimizing workflow-level performance. Similarly, Autellix \cite{luo2025autellix} adopts system-level scheduling policies that treat LLM agents as interdependent execution units. Together, Parrot and Autellix mark a shift from isolated request execution toward workflow-aware serving, where end-to-end efficiency (rather than per-call latency) becomes the optimization objective.

\noindent{\textbf{LLM Energy Consumption:}}
Several studies have examined the energy consumption and optimization of deep learning and LLM training \cite{garcia2019estimation,you2023zeus,georgiou2022green,rajput2024enhancing}.
García-Martín \etal~\cite{garcia2019estimation} surveyed energy estimation techniques in machine learning, providing a taxonomy of power models and discussing their applicability to both training and inference.
Georgiou \etal~\cite{georgiou2022green} compared the energy costs of popular deep learning frameworks (e.g., PyTorch, TensorFlow) during training, revealing notable framework-dependent variations.
FECoM~\cite{rajput2024enhancing}, a fine-grained measurement framework for TensorFlow APIs, enables API-level profiling and demonstrates how parameter size and execution time influence energy consumption. 
These works emphasize that accurate estimation, framework-level profiling, and fine-grained measurement are key enablers for energy-aware deep learning practices. \textit{Zeus} \cite{you2023zeus} developed an optimization system based on fine-grained measurements that balances performance and energy efficiency by automatically finding optimal GPU-level and job configurations for recurring deep neural network training.

Complementing training-oriented studies, a growing body of work now focuses on energy consumption during inference, which dominates cumulative usage in deployed AI systems. For example, Lahmer \etal~\cite{lahmer2022energy} and Tu \etal~\cite{tu2023unveiling} evaluated the energy usage of deep learning models on edge devices such as NVIDIA Jetson boards. Sobhani \etal~\cite{sobhani2025sustainability} conducted a systematic characterization of performance-energy trade-offs across various edge platforms (e.g., Raspberry Pi, Google Coral) for machine learning, deep learning, and LLM inference. Their analysis reveal that hardware selection, lightweight frameworks, and inference parameters collectively influence both performance and energy efficiency. 

In parallel, several software-based energy meters have emerged to enable reproducible measurement and analysis of inference workloads. Argerich and Patiño-Martínez~\cite{argerich2024measuring} propose \textit{EnergyMeter}, a profiler capable of attributing CPU, GPU, memory, and storage energy consumption during inference. Similarly, \textit{Zeus}~\cite{you2023zeus}, also supports accurate measurement of inference-time energy consumption across hardware components. Huang \etal~\cite{huang2025wattsonai} advance this direction through \textit{WattsOnAI}, which unifies energy, power, and carbon metrics into a visualization toolkit for comprehensive sustainability analysis. \textit{CodeCarbon} \cite{benoit_courty_2024_11171501} has further popularized lightweight carbon accounting by integrating runtime energy estimation directly into ML pipelines.

Finally, energy knob-based characterization and benchmarking studies systematically analyze how LLM  inference configuration parameters impact the inference energy. Samsi \etal~\cite{samsi2023words} provide one of the first cross-platform baselines, profiling the energy consumption of LLM inference across various hardware and workloads. Stojković \etal~\cite{stojkovic2024towards} identify key control parameters such as input-output length, batching, and GPU frequency scaling, which govern the performance-energy balance. Wilkins \etal~\cite{wilkins2024offline} further develop workload-based energy models that capture the joint impact of input-output lengths on runtime and energy. Maliakel \etal~\cite{maliakel2025investigating} empirically explore the interplay between task type, sequence length, and clock frequency, while Argerich and Patiño-Martínez~\cite{argerich2024measuring} demonstrated the effects of batch size, quantization, and model architecture on efficiency. Rajput \etal~\cite{rajput2025tu} show that orthogonal combinations of model-, system-, and inference-level knobs can support cascading energy savings.

\textbf{Comparison with existing work:} 
Prior research on LLM orchestration and serving systems has primarily optimized for throughput, latency, and scalability, focusing on responsiveness and resource utilization rather than energy efficiency. In contrast, studies on LLM inference energy consumption focused on accurate measurement and energy knob-based characterization, but evaluations largely remain at the single-request inference level, overlooking workflow dependencies and cross-request interactions that dominate emerging LLM applications.

Our work bridges these gaps by conducting the first comprehensive characterization of performance--energy trade-offs in multi-request LLM workflows. We systematically evaluate multiple energy knobs in representative multi-request workflows and quantify how engine-level optimizations, workflow-aware schedulers behave under interdependent requests. This unified perspective reveals new sustainability challenges such as limited batching efficiency, reduced cache reuse, and compounded critical-path latencies that are absent in single-request studies, establishing the foundation for energy-aware LLM serving and orchestration.

\section{Experimental Setup and Implementation}
\label{sec:deployment_exp_setup}

This section constructs a controlled deployment environment that integrates hardware, serving systems, measurement tools, and representative workloads. 
We begin by detailing the hardware and software stack of our testbed, followed by the deployed serving systems and the energy measurement framework. We then describe the four representative multi-request workloads that form the basis of our experiments. Together, these components provide the methodological foundation for our results in Section~\ref{sec:evaluation}, enabling a systematic assessment of how key energy knobs influence sustainability in workflow driven LLMs.

% \Idea{Fig: application ECM with python script using Zeus (ref. Kairos paper)}\\
% \Idea{Fig: Measurement workflow/block diagram}\\

\subsection{Setup}

\noindent{\textbf{Testbed Setup:}} 
To ensure consistent and reliable results, we conducted our experiments on a dedicated server (Supermicro SYS-740GP-TRT). The server is equipped with two 12-core Intel Xeon 4310 CPUs (2.1 GHz, 18 MB cache) and one NVIDIA Ampere A100 GPU \cite{choquette2021nvidia} with 40 GB of HBM2 memory. It has 128 GB of DDR4 2933 MHz RAM and a total of 26 TB of storage capacity. This single-GPU setup allows us to isolate workload-level behaviors and component-wise energy dynamics without interference from communication or synchronization overheads common in distributed settings. This isolation is essential to establish a reproducible performance--energy baseline before extending to multi-GPU scenarios.
% , comprising 3 × 8 TB SATA drives configured as RAID-5 and 2 × 960 GB SSDs configured as RAID-1. The system supports PCIe-4.0×16 interconnects. 
% Networking is provisioned through dual 10/25 GbE ports, ensuring sufficient bandwidth for distributed workloads. 
Power is supplied by redundant 2200W (80 PLUS Titanium) high-efficiency power supply units.
% , which are critical for stable GPU operation under varying power-capping settings. 
The software stack includes CUDA 12.6, cuDNN 9.1.0, and Ubuntu 24.04.2 LTS with kernel 6.5.0-27-generic.

\noindent{\textbf{Serving Systems:}} We evaluate our workloads on two serving systems, vLLM \cite{vllm} and Parrot \cite{lin2024parrot}, which embody distinct design philosophies. vLLM is widely adopted for high-throughput inference and integrates advanced optimizations such as FlashAttention \cite{dao2022flashattention}, continuous batching \cite{yu2022orca}, and PagedAttention \cite{kwon2023efficient} to improve GPU utilization and reduce memory fragmentation. We use vLLM V1 engine \cite{vllm_v1} (version 0.9.1) and access vLLM through its Python bindings, which implement an OpenAI-compatible interface for LLM engine interaction. 
In contrast, Parrot introduces semantic variables to expose inter-request dependencies, enabling workflow-aware scheduling optimizations that are particularly suited for multi-request LLM applications. This design allows Parrot to coordinate scheduling across dependent calls, targeting lower end-to-end latency for structured workloads. Parrot exposes its query submission and configuration endpoints through a FastAPI-based interface \cite{fastapi}. Together, these systems provide a representative comparison between request-level and workflow-level serving designs under performance--energy tradeoffs. All evaluations are conducted using isolated Docker containers for vLLM and Parrot. 

\noindent{\textbf{Large Language Model:}} The evaluation is based on an open-source implementation of the Llama-2 model (with 7 billion parameters), available via request from Meta, a pre-trained variant from the Llama family of decoder-only models \cite{touvron2023llama}. Decoder-only transformers have become the dominant architecture for generative LLM deployments due to their simplicity and efficiency in autoregressive token generation. Llama-2-7B, in particular, is optimized for instruction-following tasks and has been widely adopted in research as a representative open-source alternative to commercial models \cite{zheng2023judging}. The model supports a context size of 4096 tokens and is trained using conversations collected from ShareGPT \cite{shareGPT}. We use the HuggingFace implementation \cite{llama2} of Llama-2-7B with PyTorch and the Transformers library, together with the default \textit{llama-tokenizer} as the tokenizer backend. For model configuration, we set the decoding parameters at a temperature of 0.7 and a top-p value of 1.0.
% , aligning with the typical ranges reported in previous work \cite{stojkovic2024towards}. 
These settings provide a balance between diversity and determinism in generation. Moreover, the settings are consistent with the experimental setups used in both serving-system research and sustainability benchmarking \cite{stojkovic2024towards}.

\noindent{\textbf{Energy Measurement:}} 
A key enabler for energy measurement is selecting effective tools. We choose \texttt{Zeus} \cite{you2023zeus} as our energy measurement framework for LLM inference.
%In our study, we employ \texttt{Zeus} as our energy measurement framework for LLM inference. 
Zeus is explicitly designed for deep learning workloads, providing both accurate energy consumption monitoring and optimization capabilities. It integrates tools and libraries such as \texttt{nvidia-smi} \cite{nvidia-smi}, \texttt{amdsmi} \cite{amd-smi}, and \texttt{RAPL} \cite{david2010rapl}, to capture fine-grained, component-wise energy across CPU, GPU, and DRAM. This unified interface enables reproducible measurement without requiring hardware modifications, and supports a wide range of accelerators, including NVIDIA GPUs, AMD GPUs, Apple Silicon, and Jetson-class embedded platforms. Beyond measurement, Zeus also includes support for energy optimization workflows, making it an appropriate foundation for sustainability-oriented inference studies. It is widely adopted, actively maintained, and well received by the community, ensuring reliability and reproducibility for our evaluation setup. 

Zeus reports windowed average energy consumption at one-second granularity, which is well suited for capturing the dynamics of multi-request LLM workloads \cite{rajput2025tu}. All measurements are conducted under stable thermal conditions with active cooling and isolating the inference process to eliminate background interference so that only the energy consumed during model execution is captured. \Update{Listing~\ref{lst:unified-workload} presents a unified, workload-agnostic energy measurement framework that standardizes execution and metric collection across all evaluated multi-request workflows (see workload description in Section \ref{subsec:worload-spec}). In the listing, \texttt{LLMBackend} denotes an abstract serving-system interface that encapsulates the underlying LLM inference engine. In our evaluation, this abstraction is instantiated using vLLM or Parrot, enabling a uniform execution and measurement pipeline across different serving-system designs.}
% Listing~\ref{lst:agentic-coding} illustrates how we integrate Zeus to measure energy in a multi-agent coding workload (see workload description in Section \ref{subsec:worload-spec}).

\noindent{\textbf{Challenges:}} While constructing this evaluation environment, we encounter several practical challenges that required careful mitigation. First, workflow orchestrations introduce strong inter-request dependencies, making it difficult to isolate requests for batching and complicating cache management. We address this by explicitly modeling dependencies in workflow definitions and validating execution order to ensure consistency across trials. Second, concurrency control is non-trivial as heterogeneous request arrivals in multi-agent and composite workloads created variability in queueing and scheduling, which we mitigate by fixing user concurrency levels and aligning prompt templates across runs. Third, accurate energy measurement pose challenges due to noise from background processes and component-level drift. To address this, we isolate inference processes within dedicated Docker containers, stabilize thermal conditions with active cooling, and repeat each experiment ten times to capture statistically significant averages. Finally, heterogeneity across serving systems required careful configuration. For instance, vLLM’s continuous batching and Parrot’s semantic variables expose different tuning knobs, which we coordinate by standardizing decoding parameters, maximum sequence lengths, and energy measurement intervals. Following these mitigation steps, we ensure that our evaluation pipeline remained reproducible, fair, and representative of real-world LLM deployments.
% \Idea{Selecting tools to develop a measurement scheme}
% \begin{itemize}
%     \item Energy Measurement tools in data center network: software, hardware meters; add table
% \end{itemize}

\begin{lstlisting}[style=llmmini,
caption={\Update{Unified workload-agnostic pseudo-code for multi-request LLM workloads with energy measurement.}},
label={lst:unified-workload}]
from zeus.monitor import ZeusMonitor

# ... (imports, configuration, environment setup) ...

# 1) Initialize energy monitor and LLM backend
zeus_monitor = ZeusMonitor()
llm_backend = LLMBackend(...)   # abstract serving-system interface
gen_cfg = {"model": "...", "max_tokens": "...", "temperature": 0.7}

# 2) Define generic LLM inference interface
def llm_request(prompt, **gen):
    """
    Abstract LLM invocation.
    The backend encapsulates the serving system (e.g., vLLM, Parrot)
    and returns generated text along with token usage statistics.
    """
    text, token_stats = llm_backend.generate(prompt, **gen)
    return text, token_stats

# 3) Define workload-specific prompt generation
def get_workload(workload, *, chunks=None, queries=None, system_prompt=""):
    """Returns an iterable of prompts based on workload type."""
    chunks = chunks or []
    queries = queries or []

    workloads = {
        "summarize": [f"Summarize {c}" for c in chunks],
        "search":    [f"{system_prompt} Query: {q}" for q in queries],
        "agentic":   [
            "[Architect] Design...",
            "[Engineer] Implement...",
            "[Reviewer] Review...",
            "[Reviser] Revise..."
        ],
        # Composite workload: heterogeneous sub-workloads
        "composite": (
            [f"Summarize {c}" for c in chunks] +
            [f"{system_prompt} Query: {q}" for q in queries]
        ),
    }
    return workloads.get(workload, [])

# 4) Execute workload with energy and performance accounting
def run_workload(workload, *, chunks=None, queries=None, system_prompt=""):
    # Begin workflow-level energy window
    zeus_monitor.begin_window(workload)

    total_out_tokens = 0
    # Workflow-level latency is captured by the measurement window
    for prompt in get_workload(
        workload,
        chunks=chunks,
        queries=queries,
        system_prompt=system_prompt
    ):
        _, token_stats = llm_request(prompt, **gen_cfg)
        total_out_tokens += token_stats.get("output_tokens", 0)

    mes = zeus_monitor.end_window(workload)
    return mes, total_out_tokens

# 5) Report metrics
mes, out_tokens = run_workload(
    workload="workload_name",
    chunks=[...],        # used by summarize/composite
    queries=[...],       # used by search/composite
    system_prompt="..."  # used by search/composite
)
report = {
    "Energy": {
        "E_cpu (J)": mes.cpu,
        "E_gpu (J)": mes.gpu,
        "E_dram (J)": mes.dram,
        "E_total (J)": mes.total
    },
    "Latency": {
        "T_end_to_end (s)": mes.time_s
    },
    "Throughput": {
        "Tokens/s": out_tokens / mes.time_s
    },
    "Metadata": {
        "Workload": workload,
        "Model": gen_cfg["model"],
        "Gen params": gen_cfg,
        "Backend": "...",
        "Runtime env": "..."
    }
}
\end{lstlisting}

\subsection{Workloads Specification}
\label{subsec:worload-spec}
We carefully select four representative multi-request LLM workloads that reflect both practical deployment scenarios and stress-testing conditions in data center environments \cite{parrot,luo2025autellix,chen2025kairos}. Workload implementations largely follow LangChain~\cite{langchain2025}, a widely used orchestration framework for developing and deploying LLM-based applications. 

\begin{figure}[!hbtp]
\centering
\begin{subfigure}[t]{0.32\linewidth}\centering
  \adjustbox{width=\linewidth,max height=0.2\textheight,keepaspectratio,center}{%
    \includegraphics{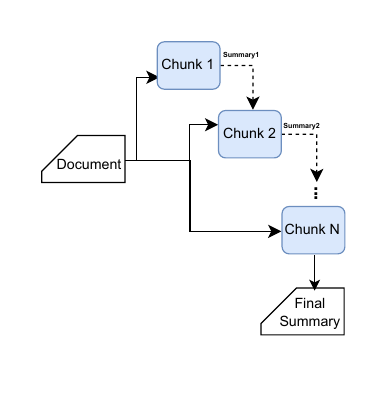}}
  \caption{}
  \label{fig:chain-summarization-workflow}
\end{subfigure}\hfill
\begin{subfigure}[t]{0.32\linewidth}\centering
  \adjustbox{width=\linewidth,max height=0.2\textheight,keepaspectratio,center}{%
    \includegraphics{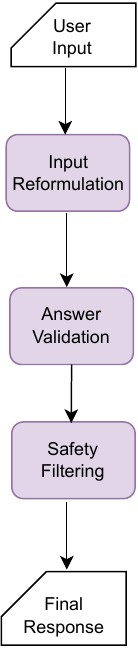}}
  \caption{}
  \label{fig:bing-copilot-workflow}
\end{subfigure}\hfill
\begin{subfigure}[t]{0.32\linewidth}\centering
  \adjustbox{width=\linewidth,max height=0.2\textheight,keepaspectratio,center}{%
    \includegraphics{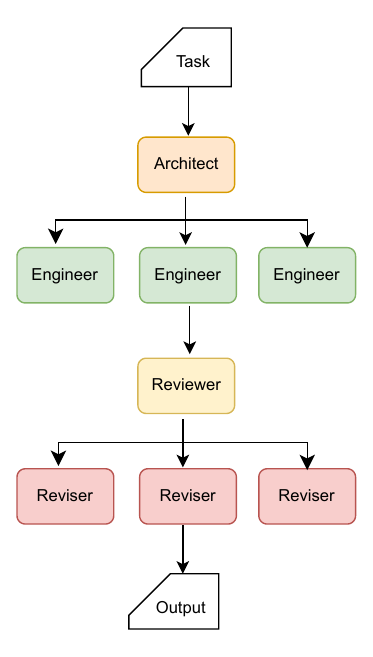}}
  \caption{}
  \label{fig:multi-agent-coding-workflow}
\end{subfigure}
\caption{Representative multi-request LLM workflows in (a) Document chain summarization, (b) LLM-powered search, and (c) Multi-agent coding workload.}
\label{fig:workflows}
\end{figure}

\noindent{\textbf{Document Chain Summarization:}}
Document chain summarization represents a key workload for knowledge-intensive domains such as scientific publishing, legal services, and government reporting, where single-pass summarization is infeasible due to the limited context length of current LLMs \cite{liu2023lost, chen2024more}. Each workflow instance typically invokes 2-40 LLM requests. To manage this multi-request dependency, the workload implements a sequential abstractive summarization process tailored for long-form documents (Figure \ref{fig:chain-summarization-workflow}).
The summarization process iteratively works on the semantically coherent segments (also known as chunks) of a source document.  
The initial summary is produced from the first segment while subsequent segments are integrated through a refinement loop based on their importance to the summary. A token budget is applied to bound the output of the evolving summary, ensuring concise and cohesive coverage of the source. In the system perspective, document chain summarization stresses serving infrastructures with strong inter-request dependencies, since the output of one LLM call is passed as part of the input to the next. As each call must generate an abstractive summary before proceeding to the next, the workload is inherently decode-heavy and sensitive to per-token generation cost. These sequential dependencies limit batching opportunities, worsen latency accumulation and introduce challenges in cache reuse and state management. 
%The source document is partitioned into semantically coherent segments, and the summarization proceeds iteratively. 
%where the current summary is updated only if new content contributes additional important information. This design enforces continuity while suppressing redundancy and drift. 

We select a large collection of academic Arxiv papers \cite{li2023unlocking} that represent long-form inputs to evaluate the effectiveness and efficiency of chain-style summarization.  Given the substantial length of many Arxiv articles, we impose a cap on the number of input tokens processed from each document, which ensures a uniform and controlled input size across evaluations, avoiding resource bottlenecks while still preserving sufficient semantic content for meaningful summarization. 
Since summarization processes large chunked inputs, it directly amplifies prefill computation, making it the only workload where input length significantly impacts energy--performance trade-offs. We therefore evaluate input length only for this workload to capture this distinctive effect. 

%%Prior studies and our measurements show that input length generally has negligible influence compared to other knobs such as batch size or power capping~\cite{chien2023reducing,wilkins2024offline}. Summarization, however, is unique in processing large, chunked inputs that directly amplify prefill computation, making it the only workload where input length materially impacts energy-performance trade-offs. By isolating this knob within summarization, we capture its distinctive effect while avoiding redundant evaluation across workloads where its impact is minimal.
%\textcolor{purple}{Given the substantial length of many Arxiv articles, we impose a cap on the number of input tokens processed from each document. This ensures a uniform and controlled input size across evaluations, avoiding resource bottlenecks while still preserving sufficient semantic content for meaningful summarization.} 
%\textcolor{purple}{In fact, this workload is also the only case where we examine the effect of input length.} 

%We utilize the Arxiv papers dataset \cite{li2023unlocking}, which contains a large collection of academic papers, offering realistic long-form inputs for evaluating the effectiveness and efficiency of chain-style summarization. 

\FloatBarrier
\noindent{\textbf{LLM-powered Search:}} This workload models an assistant-style application inspired by productivity copilots that integrate LLMs into search and office environments such as Bing Chat \cite{microsoftCopilot} and Google Gemini \cite{googleGemini}. 
% Search-oriented copilots have become one of the most common entry points for general users to interact with large language models. Consequently, even modest per-query energy costs scale to substantial aggregate consumption, making energy-performance trade-offs for this workload useful in practice \cite{iea2025efficiency, wilkins2024offline}. 
The defining characteristic of the workload is the use of a large system prompt that encodes extensive instructions, safety guidelines, and contextual knowledge before any user query is processed (Figure \ref{fig:bing-copilot-workflow}). Each interaction then appends the user’s query to this persistent prompt. Overall, each user session typically triggers 2-8 LLM calls to generate a contextually grounded response.
From an application perspective, this reflects real-world deployments, where continuous context and safety constraints must be preserved while serving diverse user requests. From a systems perspective, the workload is dominated by heavy prefill costs due to the long system prompt, followed by generation steps that require maintaining responsiveness under interactive conditions. However, unlike the document chain summarization workload, LLM-powered search maintains a static shared prefix across all requests. Consequently, its prefill cost is amortized through prefix caching, making input length variations less impactful as an energy knob and not measured separately. Overall, this combination stresses serving infrastructures by increasing memory pressure and latency in the prefill phase, while still demanding efficient throughput during response generation. 

We initialize this workload with a system prompt of approximately 3000 tokens, encoding role specifications, safety guidelines, domain-specific instructions, and few-shot exemplars. This static prefix is preserved across all user interactions, reflecting real-world deployments where safety and consistency must be enforced across sessions. On top of this shared context, we synthesize a workload of 64 requests drawn from a diverse query set (covering topics such as AI, IoT, renewable energy, and biology), each appended to the static prompt.

% \textcolor{red}{here, you are not talking about which workload or input we use and why.is not it possible adding this information in Table 2?}

\noindent{\textbf{Multi-Agent Coding:}} This workload captures collaborative software development orchestrated through multiple specialized LLM-driven agents. As in Figure \ref{fig:multi-agent-coding-workflow}, the task is framed as building a functional program (\eg a simple snake game) by decomposing the workflow into distinct roles such as architect, engineer, reviewer, and reviser. Each role contributes iteratively to the codebase where the architect specifies the system design and APIs, the programmers implement the code, the reviewers provide quality feedback, and the revisers integrate revisions based on review comments. This creates a pipeline of dependent LLM calls where intermediate outputs (\ie design specifications, code implementations, review feedback) are passed between agents to progressively refine the software artifact. The workflow typically comprises 15 to 20 LLM calls across all the agents. From a systems perspective, the workload stresses serving infrastructures with dynamic, role-driven workflows that consist of multiple interdependent requests. The execution graph exhibits a directed acyclic graph (DAG) where downstream tasks cannot proceed until upstream agents complete, limiting opportunities for batching and introducing sensitivity to latency accumulation. To align with emerging agentic frameworks, our implementation follows the MetaGPT paradigm \cite{hong2023metagpt}, which provides structured role assignments for multi-agent collaboration. 

In our implementation, each agent role is expressed as a prompt template, and their interactions form a fixed DAG of dependent calls. Input and output token limits are capped per role (e.g., $\sim$1000 tokens for coding, $\sim$500 for review/revision) to ensure controllable execution while preserving semantic fidelity. This setup reflects practical multi-agent deployments and provides consistent measurement across repeated trials \cite{chaudhry2025murakkab}.

\noindent{\textbf{Composite Workload:}} 
% This workload integrates heterogeneous applications to capture the diversity of LLM usage in practical deployments.
This workload integrates heterogeneous applications to capture the diversity of LLM usage in practical deployments. By doing so, it also mimics datacenter multi-tenant scenarios where diverse services share the same GPU pool, allowing us to study interference effects and resource contention patterns.
For implementation, we combine the chain summarization workload with the LLM-powered search workload and execute them concurrently on the same vLLM/Parrot deployment (with energy monitoring) to capture realistic multi-application contention. While the LLM-powered search application is characterized by long static system prompts and short, interactive user queries, the chain summarization task involves multi-stage iterative refinement over long-form documents with strong inter-request dependencies. Running these workloads concurrently creates a composite scenario where distinct execution patterns coexist. Prefix-heavy, latency-sensitive copilot queries overlap with sequential, stateful summarization pipelines where each workflow instance issues a variable number of LLM calls drawn from both the summarization and search pipelines. From a systems perspective, such mixed settings stress serving infrastructures with competing demands on batching efficiency, cache reuse, and scheduling fairness. The heterogeneity amplifies contention for GPU resources, as workloads vary in prompt length, output size, and interactivity.
% \textcolor{red}{I feel you describe a lot about the workload but very little about their implementation or deployment details relevant to the measurement. check it again please. BTW, where in the text you refer Fig 2?}

% \textcolor{red}{should not it go in the evaluation section to justify why we have done this? --- Although our study explores the impact of multiple energy knobs, we specifically evaluate the effect of input length only within the document chain summarization workload. Prior literature and our own measurements confirm that, for most applications, input length has negligible effect compared to other energy knobs on inference energy consumption~\cite{wilkins2024offline,chien2023reducing}. Document summarization, however, involves processing large, chunked inputs that directly amplify prefill computation, making it the only workload where input length materially influences performance-energy trade-offs. By isolating this knob in the summarization case, we capture its unique impact while avoiding redundant evaluation across workloads where its effect is minimal.}
% \textcolor{red}{implementation??}

\subsection{Metrics}
\label{subsec:EM}
We evaluate the following metrics, repeating each experiment ten times and reporting the mean with a 95\% confidence interval.
%\textcolor{purple}{, confirming that the observed patterns are consistent and not artifacts of transient load or thermal variance.}
% \textcolor{red}{are these standard metrics? did you define them following standard definitions? any supportive references? how about other energy measurement works? do they use the same metrics?}

\noindent{\textbf{Workflow Mean Latency:}} For a given workflow, the end-to-end latency is measured as the elapsed time between the submission of the first request and the completion of the final request in the workflow. The overall metric is reported as the arithmetic mean across all runs in the experiment.
Unlike per-request metrics such as Time-to-First-Token (TFTT) or Time-per-Output-Token (TPOT), workflow-level latency reflects the holistic performance of multi-request applications, making it a more representative measure for complex LLM-based workloads \cite{parrot, luo2025autellix}, which we report in seconds. For brevity, we refer this metric as latency throughout the rest of the paper.

\noindent{\textbf{Throughput:}} We report throughput in tokens per second (tokens/s), which we compute over the generation phase only, as this phase dominates runtime and energy and best reflects steady-state efficiency. Throughput quantifies the total number of tokens generated per unit time across completed requests, capturing how effectively the system sustains token production. It serves as a key metric for analyzing performance-energy trade-offs in LLM inference, complementing latency-based measures ~\cite{samsi2023words, chitty2024llm}.

\noindent{\textbf{Workflow Mean Total Energy:}}
We measure the total energy consumed by each workflow by aggregating per-component power traces (CPU, GPU, and DRAM) over the workflow’s execution interval. The total energy for each run is computed by summing the sampled power readings across all components and time windows, using RAPL for CPU/DRAM and NVML for NVIDIA GPU monitoring. The workflow mean total energy is then obtained as the average across all experiment runs. We report values in joules (J). This metric captures the full-system energy cost across major hardware components, providing a direct measure of sustainability impact beyond what latency or throughput alone can represent~\cite{stojkovic2024towards, argerich2024measuring}. For brevity, we would refer to this metric as total energy.

\noindent{\textbf{Energy per Token:}} We measure energy efficiency using the metric of energy per token, defined as the ratio between the total energy consumed during the workflow execution and the total number of output tokens generated. This metric normalizes energy consumption by the volume of useful work performed, capturing how efficiently the serving system converts energy into generated tokens \cite{samsi2023words}. We report results in joules per token (J/token).
\section{Results}
\label{sec:evaluation}
This section presents the evaluation results by answering the following research questions. 

%To guide our evaluation, we address the following research questions:
\begin{itemize}
        \item[RQ1:] What workload- and system-level energy knob(s) most significantly influence the performance and efficiency of multi-request LLM workloads?
        \item[RQ2:] How does the most influential  energy knob (identified in RQ1) shape the energy-performance trade-offs across representative multi-request LLM workloads?
        \item[RQ3:] For the most energy-demanding workload, how do state-of-the-art serving systems (vLLM and Parrot) compare in terms of performance and energy efficiency?
\end{itemize}

\subsubsection{RQ1: Impact of energy knobs in multi-request LLM applications}
\label{subsubsec:rq1}
To answer the first research question, we begin by considering three representative knobs that are widely exposed in serving systems: output length, batch size, and GPU power capping. These knobs reflect common configuration levers available to practitioners and they allow us to explore how different workload- and system-level choices affect both performance and energy consumption. We evaluate their impact using the \textit{LLM-powered search} workload from our previously defined set of representative applications. We select this workload to answer RQ1 because search-oriented copilots have become one of the most common entry points for general users to interact with large language models. Consequently, even modest per-query energy costs can scale to substantial aggregated consumption, making performance--energy trade-offs in this workload especially impactful in practice \cite{iea2025efficiency,wilkins2024offline}.
% (see Section~\ref{subsec:worload-spec}). 
% This workload, which generalizes the category of interactive copilots, is particularly relevant as it mirrors widely adopted deployments such as Bing Chat \cite{microsoftCopilot} and Google Gemini \cite{googleGemini}. Its importance stems from the magnitude of usage: search-oriented copilots have become one of the most common entry points for general users to interact with large language models. Consequently, even modest per-query energy costs scale to substantial aggregate consumption, making energy-performance trade-offs for this workload especially impactful in practice \cite{iea2025efficiency, wilkins2024offline}.

\begin{figure}[!htbp]
  \centering
  \begin{subfigure}{0.32\columnwidth}
    \centering
    \includegraphics[width=\linewidth]{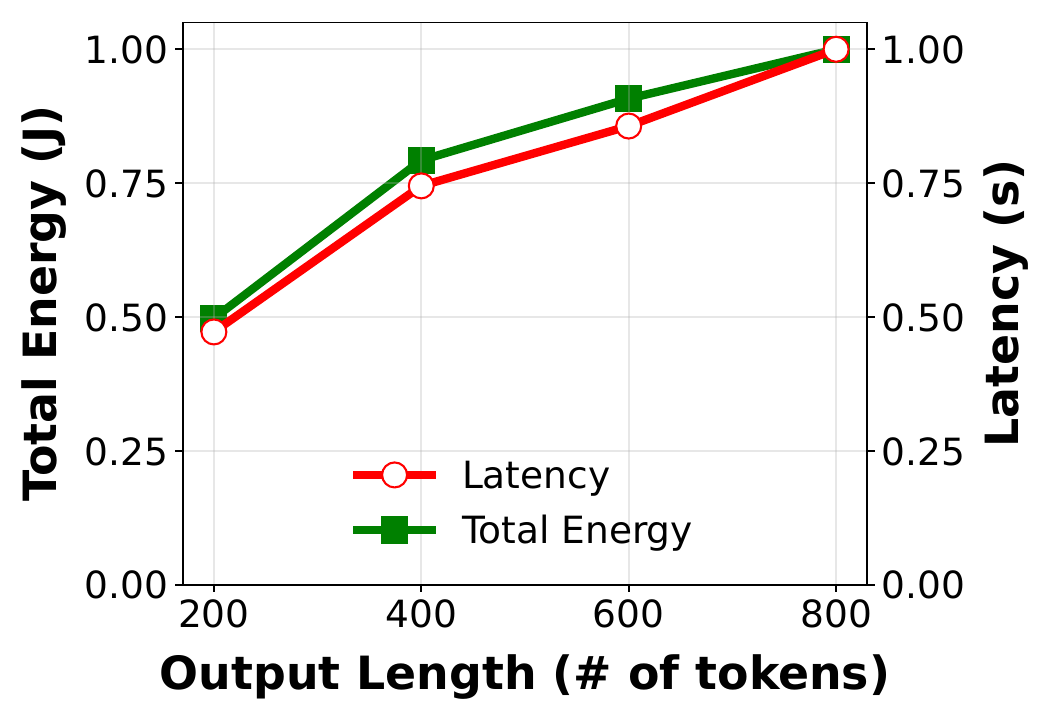}
    % \caption{Energy/Latency vs. Output Length}
    \caption{}
    \label{fig:rq1-outsize-lat-eg}
  \end{subfigure}
  \begin{subfigure}{0.32\columnwidth}
    \centering
    \includegraphics[width=\linewidth]{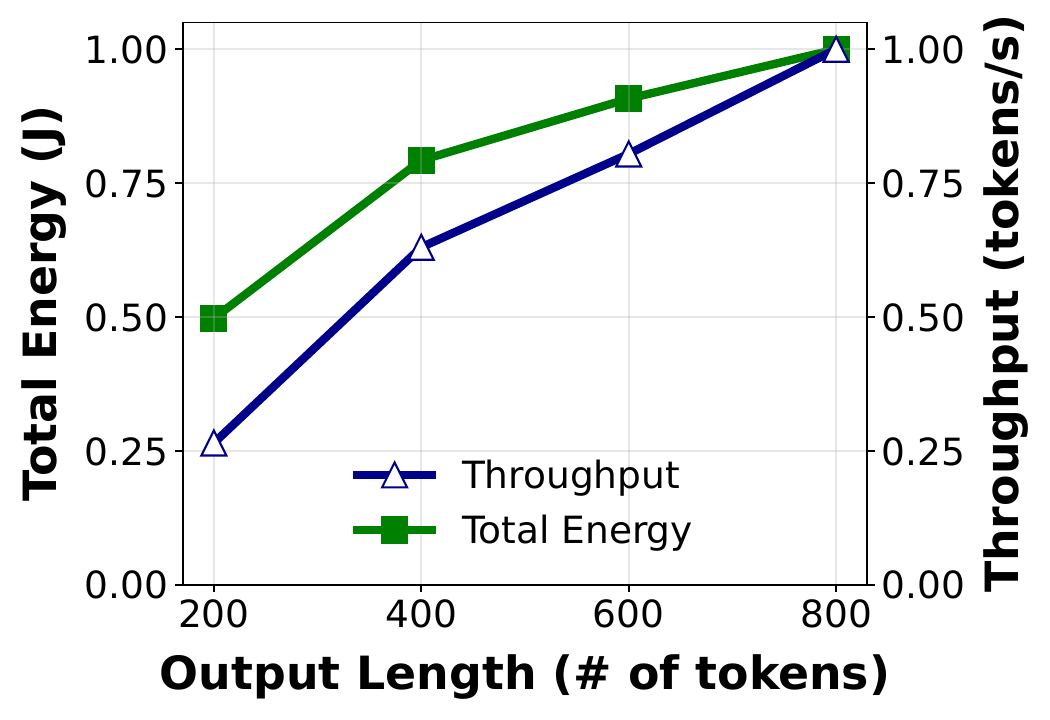}
    % \caption{Energy/Throughput vs. Output Length}
    \caption{}
    \label{fig:rq1-outsize-tp-eg}
  \end{subfigure}
  % \hfill
  \begin{subfigure}{0.32\columnwidth}
    \centering
    \includegraphics[width=\linewidth]{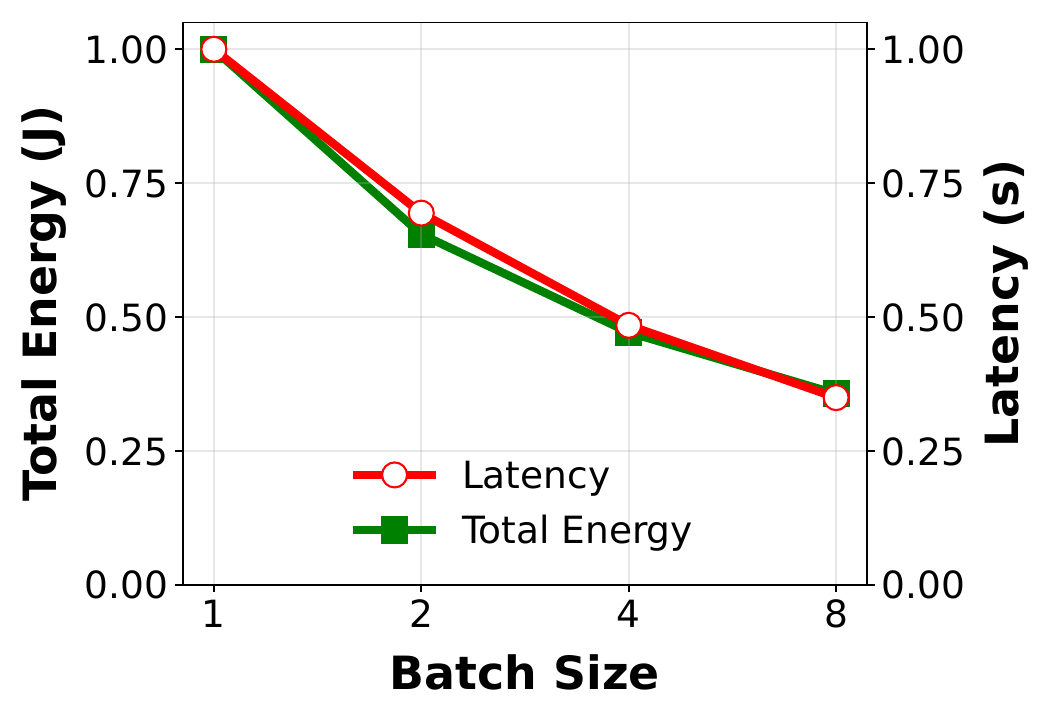}
    % \caption{Energy/Latency vs. Batch Size}
    \caption{}
    \label{fig:rq1-batch-lat-eg}
  \end{subfigure}
  % \vspace{4pt}
  \begin{subfigure}{0.32\columnwidth}
    \centering
    \includegraphics[width=1\linewidth]{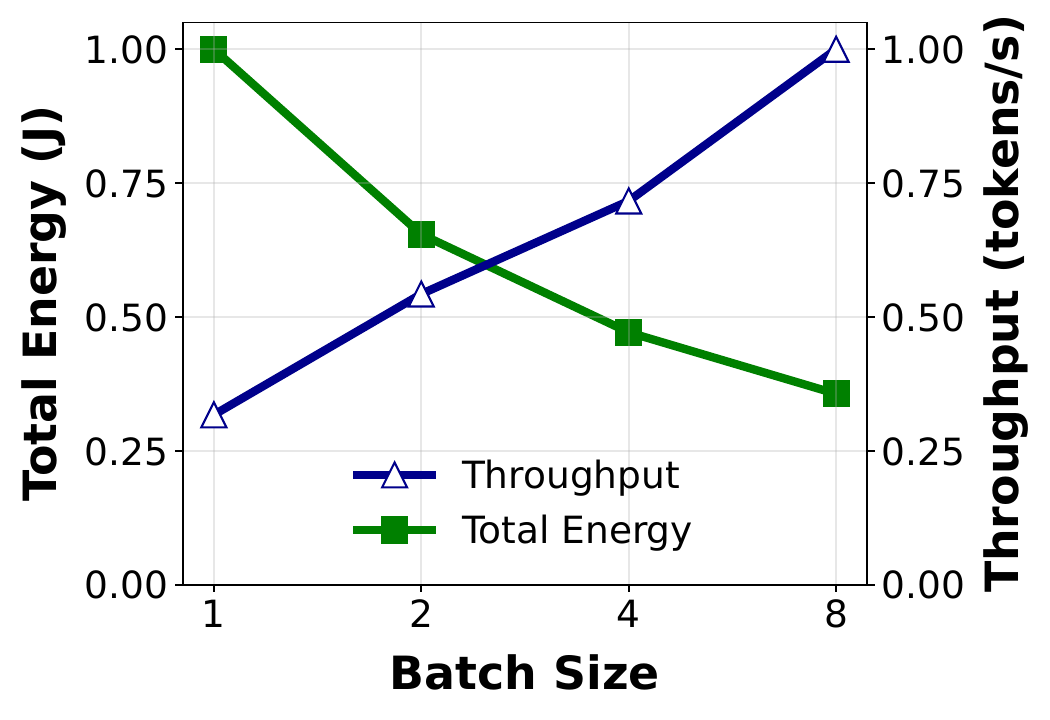}
    % \caption{Energy/Throughput vs. Batch Size}
    \caption{}
    \label{fig:rq1-batch-tp-eg}
  \end{subfigure}
  \begin{subfigure}{0.32\columnwidth}
    \centering
    \includegraphics[width=\linewidth]{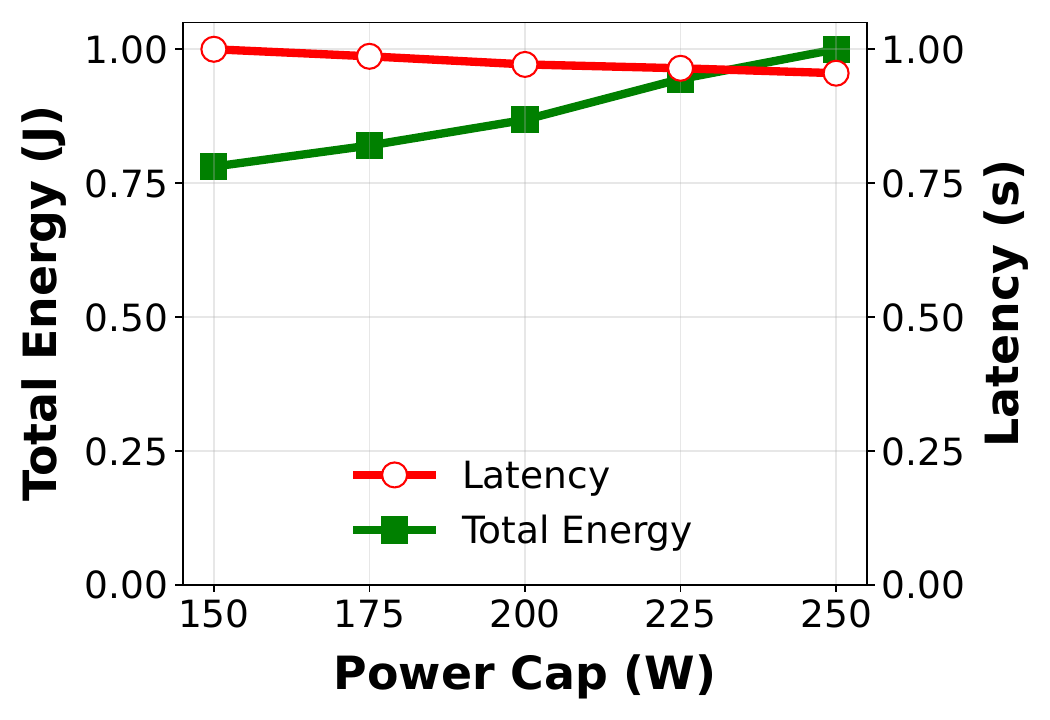}
    % \caption{Energy/Latency vs. Power Cap}
    \caption{}
    \label{fig:rq1-pcap-lat-eg}
  \end{subfigure}
  \begin{subfigure}{0.32\columnwidth}
    \centering
    \includegraphics[width=\linewidth]{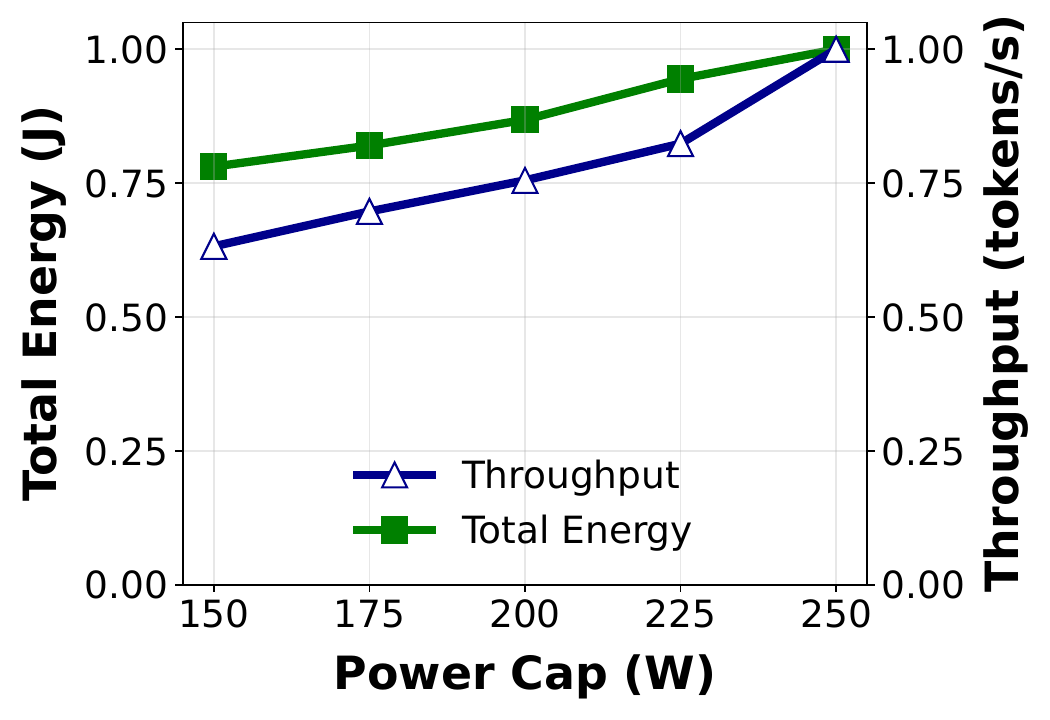}
    % \caption{Energy/Throughput vs. Power Cap}
    \caption{}
    \label{fig:rq1-pcap-tp-eg}
  \end{subfigure}
    \caption{RQ1 results on the \textit{LLM-powered search} workload. Panels (a-b) sweep output length, (c-d) sweep batch size, and (e-f) sweep GPU power cap, with all other knobs fixed. Each panel uses dual Y-axes with normalized total energy in left vs. normalized performance (latency/throughput) in right Y axis.}
  % \caption{RQ1 results on the \textit{LLM-powered search} workload. Each row varies one knob (output length, batch size, power cap) while holding the others at default values. Plots use dual Y-axes to show normalized performance (latency/throughput) against normalized energy.}
  \label{fig:rq1-sixplots}
\end{figure}

From an implementation standpoint, 
% the application is initialized with a system prompt of approximately 3000 tokens, encoding role specifications, safety guidelines, domain-specific instructions, and few-shot exemplars. This static prefix is preserved across all user interactions, reflecting real-world deployments where safety and consistency must be enforced across sessions. On top of this shared context, we synthesize a workload of 64 requests drawn from a diverse query set (covering topics such as AI, IoT, renewable energy, and biology), each appended to the static prompt. 
we set \texttt{enable\_prefix\_caching=True} to share static prefixes across concurrent requests. Responses are capped at a maximum of 800 output tokens, aligning with typical constraints in interactive copilots where latency and readability are prioritized over arbitrarily long generations. To capture concurrency effects, the workload is executed with 8 simultaneous users, each issuing separate independent queries in parallel. This design stresses both the prefill and generation phases of LLM inference. Prefill latency is dominated by the long static prompt, while generation must sustain responsiveness under concurrent load. We conduct this experiment using the vLLM serving system as it has emerged as the de facto in the domain of open-source LLM serving due to its high-throughput design, advanced batching strategies, and efficient memory management mechanisms. Figure~\ref{fig:rq1-outsize-lat-eg}-\ref{fig:rq1-pcap-tp-eg} depict the effect of varying one of the knobs (output length, batch size, and GPU power cap) while keeping the other two fixed at their default settings, illustrating the corresponding latency--energy or throughput--energy relationships. Here, we report normalized values for latency, throughput, and energy to facilitate consistent scaling and comparability across knobs. 
% \textcolor{purple}{For each metric, we first compute the mean across ten repeated runs for a given configuration and then normalize these mean values by the maximum observed mean within that workload, ensuring consistent scaling and comparability.}
% Each row corresponds to one knob, with latency-energy normalization plot is on the left and throughput-energy normalization on the right. 

\FloatBarrier
Figure~\ref{fig:rq1-outsize-lat-eg} shows the effect of varying output length (200-800 tokens) where increasing the output length significantly raises latency and energy consumption. Increasing the output length leads to a near-linear rise in both workflow latency and energy consumption. This trend is expected since the generation phase dominates the inference cost. Each additional token requires a full forward pass through the model decoder stack. The gap between latency and energy remains tight, indicating that energy scales proportionally with time, as longer generations sustain GPU activity at high utilization. In Figure~\ref{fig:rq1-outsize-tp-eg}, throughput increases with output length because longer sequences produce more generated tokens per request. However, the energy cost grows at a comparable rate, resulting in diminishing efficiency improvements. In other words, while raw throughput benefits from longer outputs, the energy per token does not improve significantly.
Increasing the batch size (1-8) reduces both latency and energy consumption (Figure~\ref{fig:rq1-batch-lat-eg}). The sharpest drop occurs when moving from batch size 1 to 2, reflecting the substantial efficiency gains from amortizing prefill and kernel launch overhead across multiple concurrent requests. Beyond this point, the reductions continue but with progressively smaller gains. Importantly, the energy profile improves consistently with larger batches, demonstrating that batching is one of the most effective levers to reduce per-workflow energy. As depicted in Figure~\ref{fig:rq1-batch-tp-eg}, throughput increases steadily with larger batches, while energy consumption drops. So, batching offers twofold benefits: it increases the number of tokens processed per unit of energy while reducing the total energy expenditure. 
% However, larger batches can also introduce queuing delays, which may negatively impact interactivity in real deployments. This creates a trade-off between maximizing efficiency and ensuring responsiveness.

Figure~\ref{fig:rq1-pcap-lat-eg} shows that 
% reducing the power cap from 250W (system default) to 150W yields a moderate decrease in total energy, accompanied by a slight increase in latency. 
\Update{increasing the GPU power cap from 150W to 250W (system default) yields a moderate increse in total energy and does not monotonically reduce latency.}
At the lowest cap (150W), the system operates at significantly reduced energy levels, but latency penalties become noticeable. As the cap is raised toward 200W, latency improves while energy remains relatively controlled, suggesting an efficiency sweet spot. 
% Beyond 200W, total energy increases with little to no latency improvement, indicating diminishing energy-efficiency returns from higher power budgets. 
\Update{Increasing the GPU power cap beyond approximately 200W does not reduce latency and instead leads to a slight increase. This behavior indicates that the LLM-powered search workload is no longer power-limited in this regime, as both prefill and decode phases already operate near sustained high GPU utilization. Once the critical path is saturated, additional power does not translate into shorter execution time. Instead, we suspect that latency variations in this regime are dominated by secondary system-level effects such as dynamic voltage and frequency scaling (DVFS)\footnote{\Update{
DVFS refers to hardware mechanisms that dynamically adjust processor voltage and clock frequency to balance performance and power consumption.}} state transitions and transient scheduling noise\footnote{\Update{Transient scheduling noise refers to short-term variability in execution timing caused by runtime scheduling decisions, such as request admission, batching, and kernel dispatch, that introduce non-deterministic delays.}} introduced by continuous batching. Consequently, higher power caps primarily increase instantaneous power draw without improving end-to-end latency, explaining the non-monotonic latency trend at higher power levels.}
As seen in Figure~\ref{fig:rq1-pcap-tp-eg}, throughput increases with power cap but at a roughly proportional increase in energy, suggesting limited benefit beyond moderate cap settings. Overall, this suggests an optimal region ($\sim$200W) for balancing efficiency and responsiveness. 
% This aligns with previous findings on power capping-based energy tuning in inference workloads \cite{stojkovic2024towards, wilkins2024offline}.
% Table~\ref{tab:impactful-eg-knob-comparison} summarizes the impact of the energy knobs in LLM-powered search application workload.

% \begin{table}[!hbtp]
%     \centering
%     \begin{tabularx}{\textwidth}{|l|l|l|X|}
%     \hline
%     \textbf{Knob} & \textbf{Latency Impact} & \textbf{Throughput Impact} & \textbf{Total Energy Impact} \\
%     \hline
%     \textbf{Output length} & Positive \& Near-Linear & Positive \& Near-Linear & Positive \& Near-Linear \\
%     \hline
%     \textbf{Batch Size} & Inverse \& Non-Linear & Strong Positive \& Non-Linear & Inverse \& Non-Linear \\
%     \hline
%     \textbf{Power Cap} & Inverse \& Non-Linear & Positive \& Near-Linear & Positive \& Near-Linear \\
%     \hline
%     \end{tabularx}
%     \caption{Comparative analysis of energy knobs in LLM-based search workload.}
%     \label{tab:impactful-eg-knob-comparison}
% \end{table}
\findingsbox{We observe clear performance--energy patterns across the three knobs. Batch size is the most impactful, giving strong throughput gains and lower total energy. Output length increases energy use almost linearly, while power capping works mainly as a fine-tuning control under power limits. Overall, there is a potential for hierarchical energy optimization scheme prioritizing batch size optimization for a given workload following power capping assisted fine-tuning energy usage. Unlike single-request inference studies where these energy knobs show uniform, monotonic improvements in efficiency \cite{stojkovic2024towards,samsi2023words}, our LLM-powered search results reveal that shared-prompt reuse amplifies batching gains. This indicates that the same knobs behave in different magnitude once cross-request context sharing and concurrency are introduced, reshaping the efficiency-latency trade-off beyond single-request settings.}

% \textcolor{red}{we may remove this --- For developers aiming for sustainable deployment of multi-request LLM workflows, the optimization strategy should follow a clear hierarchy: prioritize finding the optimal batch size for the workload, then use power capping for fine-tuning. This multi-level approach is essential for achieving both responsiveness and energy efficiency in practice. }
\subsubsection{RQ2: Cross-workload effects of batch size}
\label{subsubsec:rq2}

\begin{figure}[!hbtp]
  \centering
  %==================== Subfigure A ====================
  \begin{subfigure}{0.48\linewidth}
    \centering
    \includegraphics[width=\linewidth,height=0.3\textheight,keepaspectratio]{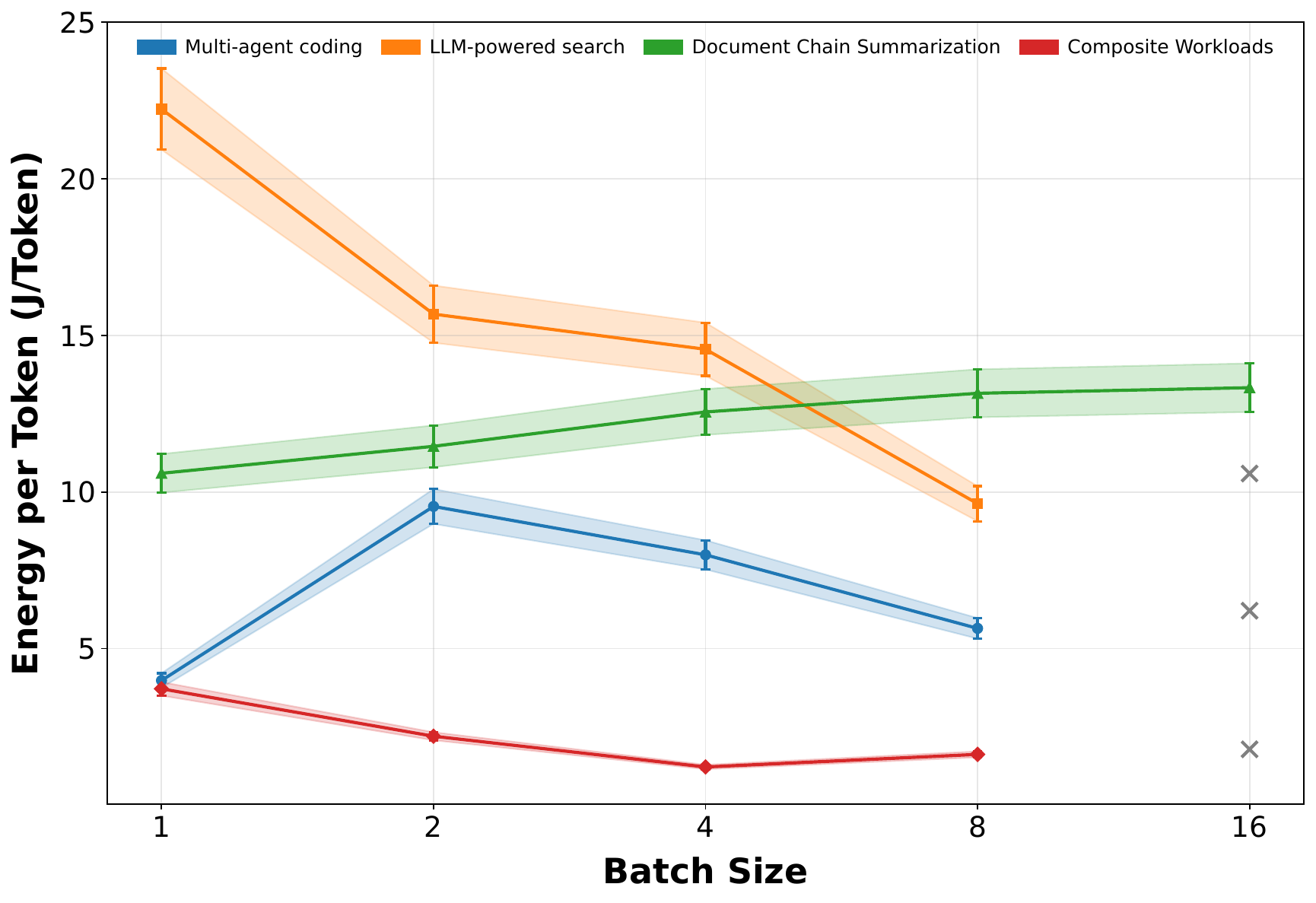}
    % \caption{Energy per token vs. batch size under 16 concurrent users. Lines show means; shaded regions are 95\% confidence intervals. Lower is better. The X markers denote batch sizes that were infeasible (OOM).}
    \caption{}
    \label{fig:rq2-bsz-energy}
  \end{subfigure}\hfill
  %==================== Subfigure B ====================
  \begin{subfigure}{0.48\linewidth}
    \centering
    \includegraphics[width=\linewidth,height=0.3\textheight,keepaspectratio]{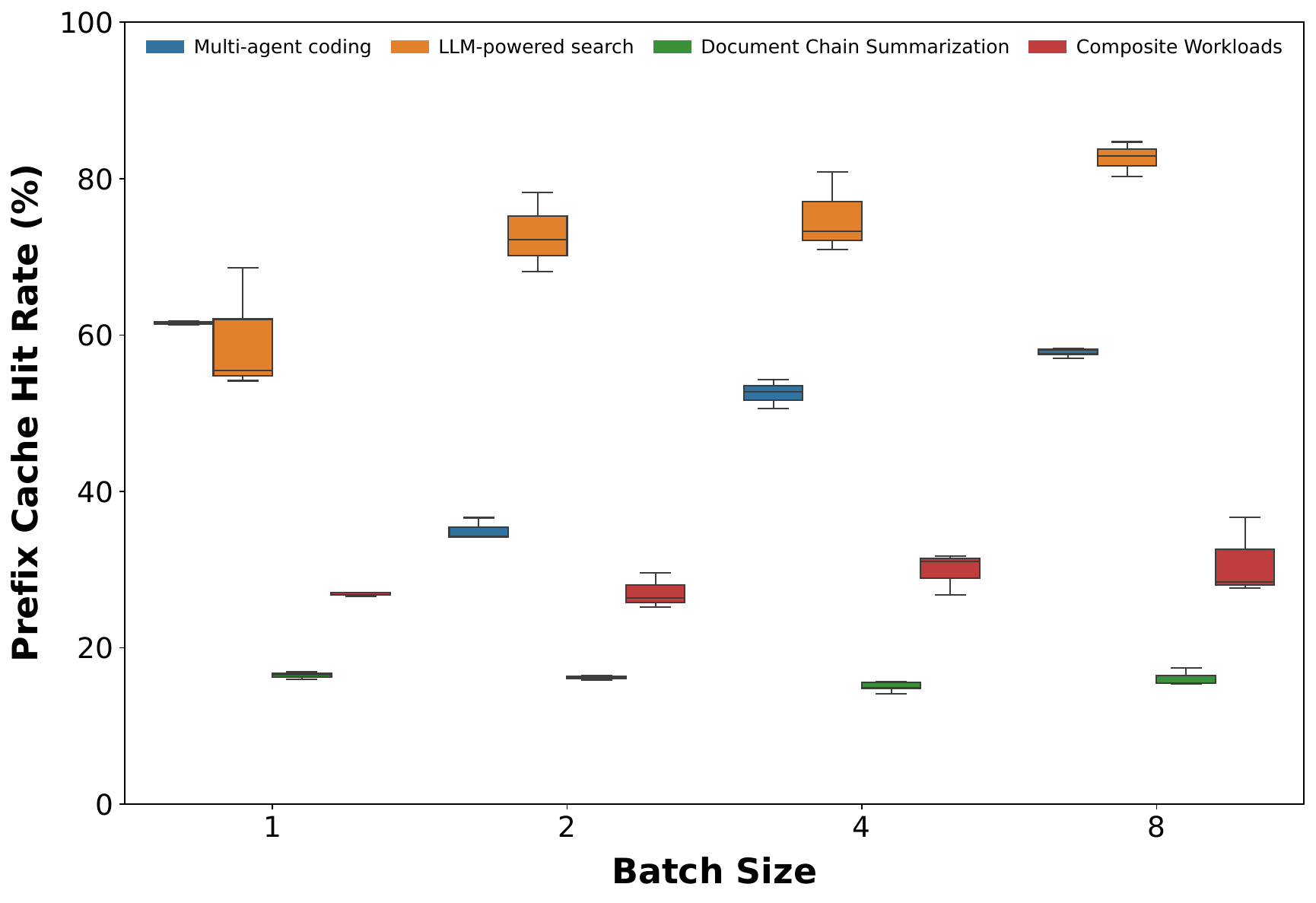}
    % \caption{Prefix cache hit rate distribution across workloads and batch sizes (1--8). Boxplots show medians and variability.}
    \caption{}
    \label{fig:rq2-prefix-cache-boxplot}
  \end{subfigure}

  % \caption{Effect of batch size on energy per token and prefix cache hit rate across workloads.}
  \caption{\emph{a)} Energy per token vs. batch size under 16 concurrent users. Lines show mean value; shaded regions are 95\% confidence intervals. Lower is better. The X markers denote batch sizes that are infeasible (OOM). 
  % \Update{The spike observed for multi-agent coding at batch size 2 is caused by heterogeneous agent-role requests being co-batched, which limits prefix-cache reuse while incurring batching and scheduling overheads.}
 \emph{b)} Prefix cache hit rate distribution across workloads and batch sizes (1--8). Boxplots show medians and variability.}
  \label{fig:rq2}
\end{figure}

To evaluate how the most impactful energy knob (batch size, as identified in RQ1) influences multi-request workloads, we sweep the batch size parameter for all four workloads introduced in Section~\ref{subsec:worload-spec}. We fix concurrency at 16 users per workload to ensure comparability and use workload-specific output lengths (maximum completion tokens) consistent with real deployments. All experiments are conducted on \emph{vLLM} with prefix caching enabled. Figure~\ref{fig:rq2-bsz-energy} presents the energy per token (J/Token) for all workloads across batch sizes 1--16 (as feasible on our NVIDIA A100-40GB GPU). Here, we plot the mean across repeated runs; the bands show 95\% confidence intervals capturing run-to-run variability (we use the same run budget and aggregation method as in Section~\ref{subsubsec:rq1}). If a configuration leads to GPU out-of-memory (OOM) due to the massive KV cache stored, we record it as infeasible and do not report a measured point. 
% Distinct workload-specific trends emerge. 

In document chain summarization workload, energy per token increases monotonically with batch size, showing that this decode-heavy workload does not benefit from batching. In contrast, LLM-powered search exhibits the opposite pattern where energy per token drops significantly as batch size increases, reaching its minimum at batch size~8 (corresponding to a $2.6\times$ improvement in efficiency). This outcome is explained by the large shared system prompt (approximately 3000 tokens shared across all requests) which enables high prefix-cache reuse across concurrent queries. 
% Multi-agent coding shows more complex behavior as the lowest energy per token occurs at batch size~1, with efficiency degrading at larger batches but partially recovering at batch size~8. Notably, batch size~2 appears as an outlier with high energy per token. This anomaly arises from scheduling inefficiencies of role-specific prompts (e.g., architect, coder, reviewer) which vary widely in structure. As a result, batching two such requests together prevents prefix reuse while still incurring queuing costs, inflating GPU energy without corresponding throughput gains. 
% ====
\Update{Multi-agent coding shows more complex behavior as the lowest energy per token occurs at batch size~1, with efficiency degrading at larger batches but partially recovering at batch size~8. Notably, batch size~2 appears as an outlier with high energy per token. This non-monotonic behavior is primarily driven by disproportionate execution characteristics across agent roles and their interaction with request scheduling in the multi-agent DAG, rather
than from measurement noise or run-to-run variance. At batch size~2, the serving system frequently co-batches requests originating from different agent roles (\eg architect, engineer, reviewer), whose computational profiles differ
substantially in terms of prompt length, decode intensity, and runtime. These roles use structurally distinct prompt templates with minimal shared prefixes, leading to poor KV-cache reuse during the prefill stage. As a result, the system incurs batching-related overheads such as queuing delay, GPU residency, and kernel launch costs, without achieving the amortization benefits typically associated with batching. This imbalance inflates GPU energy consumption while resulting in little or no throughput improvement, manifesting as a sharp increase in energy per token. This effect diminishes at larger batch sizes (\eg batch size 8), where requests from homogeneous roles are more likely to be grouped together. Such batching improves prefix alignment and cache reuse, allowing overheads to be amortized more effectively across requests. Consequently, the energy efficiency partially recovers at higher batch sizes despite increased concurrency. This behavior highlights that, in multi-agent workflows, batching efficiency depends not only on batch size but also on the semantic and structural alignment of co-batched requests.}
% ====
Finally, for the composite workload, efficiency improves up to batch size~4 due to moderate reuse, but shows reduced marginal benefits thereafter. Overall, sequential dependencies in summarization, role heterogeneity in coding, and mixed patterns in the composite workload restrict batching benefits, while LLM-powered search consistently gains from shared-prefix caching.

To further support this analysis, we leverage \emph{vLLM}’s prefix cache hit rate, defined as the fraction of prefill tokens whose key-value states are reused from prior requests rather than recomputed. A higher hit rate directly translates to reduced redundant computation, lower latency, and improved energy efficiency. Figure~\ref{fig:rq2-prefix-cache-boxplot} shows the distribution of prefix cache hit rates across batch sizes, excluding batch size~16, which is infeasible for most workloads due to GPU memory limits. The results reveal that LLM-powered search achieves both high hit rates and high variability across runs. This variability arises because, although all queries share the same static prompt, appended user queries differ in length and structure. When batch formation aligns queries with similar token lengths, reuse is maximized; when lengths diverge, reuse drops, producing variability. Multi-agent coding maintains moderate and stable hit rates ($\sim$45-60\%) across batch sizes, reflecting partial reuse from recurring role templates (e.g., coder and reviewer prompts) despite structural diversity among agents. The composite workload exhibits intermediate hit rates ($\sim$25-35\%) due to the interplay between LLM-powered search and summarization phases where search queries enable reuse, summarization stages contribute less shared context. In contrast, document chain summarization shows consistently low hit rates, as its sequential dependencies and evolving prompts limit prefix overlap across requests. Overall, the prefix cache analysis confirms our earlier observations. Workloads with large, stable shared prefixes (e.g., search) derive the most benefit from batching, whereas sequential or heterogeneous workflows exhibit constrained opportunities for reuse.

\findingsbox{The impact of batch size on energy efficiency varies notably between single- and multi-request inference. In single-request studies, larger batches almost uniformly improve throughput and reduce per-token energy by reducing computation \cite{chitty2024llm,maliakel2025investigating}. In contrast, our multi-request evaluation reveals irregular and workload-specific behavior. While LLM-powered search benefits substantially from prefix reuse (up to $2.6\times$ energy reduction), sequential or heterogeneous workflows such as summarization and multi-agent coding show reverse trends where batching increases latency and energy due to queuing delays and fragmented KV caches. This highlights a key difference from prior work; the efficiency of batching in multi-request workflows is driven not only by GPU occupancy \cite{fernandez-etal-2025-energy}, but also by cross-request dependency structure and cache reuse dynamics.}

% \findingsbox{The impact of batch size on energy usage is workload-dependent. LLM-powered search consistently benefits from larger batches due to extensive prefix reuse, while summarization and multi-agent coding see limited or even negative gains because of queuing delays and heterogeneous prompts. Composite workloads fall in between, balancing moderate reuse with overhead. The prefix cache analysis confirms that only workloads with a large, stable shared context achieve consistently high reuse, whereas others remain constrained by workflow structure. \textcolor{purple}{These patterns broadly align with single-request studies where larger batches reduce per-token energy by amortizing computation \cite{fernandez-etal-2025-energy}. But multi-request workflows amplify variability as inter-request dependencies and asynchronous arrivals introduce queuing and cache-fragmentation effects absent in single calls, sometimes reversing the efficiency trend. Hence, batch-size tuning must be application-aware; uniformly increasing batch size may degrade efficiency for certain workflows while delivering substantial improvements for others.}} 
% \textcolor{red}{agian, these results how much match with previously reported ones for single-resquest task? anything new and interesting for multi-request tasks?}
%To conclude, our cross-workload evaluation of batch size reveals that its impact on energy efficiency is highly workload-dependent. 

\subsubsection{RQ3: Comparisons of state-of-the-art serving systems} 
% Comparative analysis of serving systems for chain summarization workload}
\label{subsubsec:rq3}
To answer RQ3, we experiment with the document chain summarization workload, identified in RQ2 as the most energy-intensive application with consistently high energy per token. This workload represents long-form document processing pipelines, where sequential summarization stages accumulate both latency and energy costs. Its decode-heavy nature and limited prefix-cache reuse make it a stringent test case for evaluating the sustainability of serving systems. 
As part of this evaluation, we randomly selected ten long-form documents from the Arxiv dataset, each containing more than 36,000 tokens. These extensive inputs serve as a realistic stress test for serving systems, as their size far exceeds the context window of the LLM under test and necessitates chunk-based processing. 

We perform experiments using both \texttt{vLLM} and \texttt{Parrot} serving systems, systematically varying our key energy knobs. For each knob, we report three evaluation metrics: latency (s), throughput (tokens/s) and total energy (J) broken down across CPU, GPU, and DRAM. To ensure fairness, both serving systems are configured with identical decoding parameters, prompt templates, and measurement intervals.
% This holistic view highlights how serving system design interacts with workload-specific properties under critical energy-performance trade-offs.

\begin{figure}[!hbtp]
    \centering
    % Row 1: Chunk Size
    \begin{subfigure}{0.32\columnwidth}
        \includegraphics[width=\linewidth]{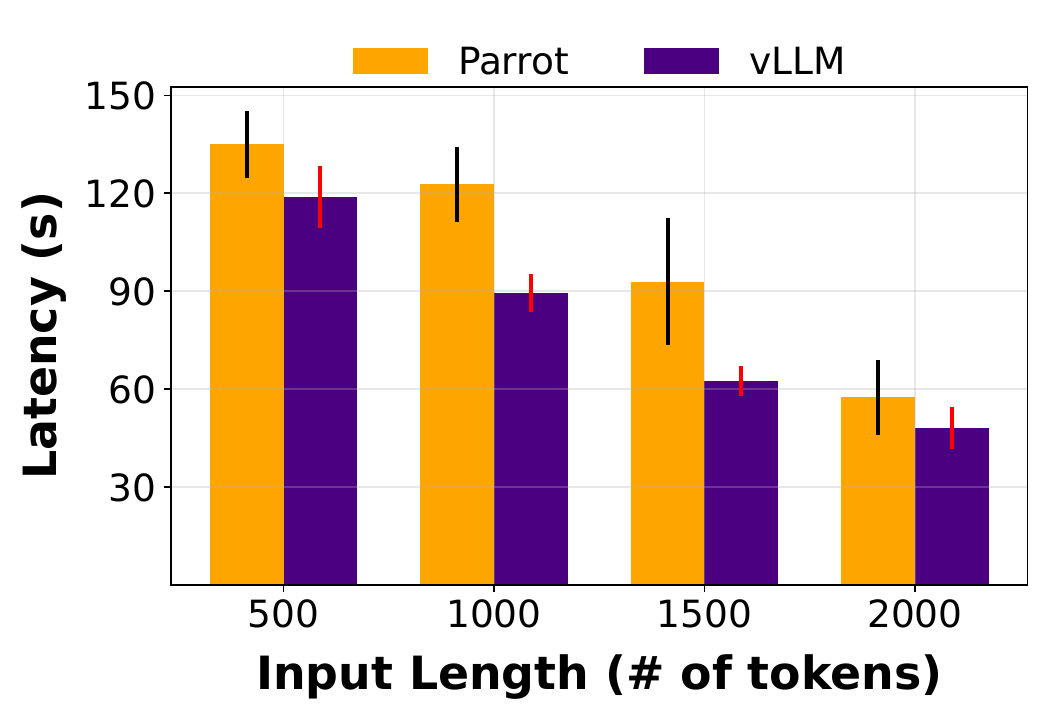}
        % \caption{Latency vs. Chunk Size}
    \end{subfigure}
    \hfill
    \begin{subfigure}{0.32\columnwidth}
        \includegraphics[width=\linewidth]{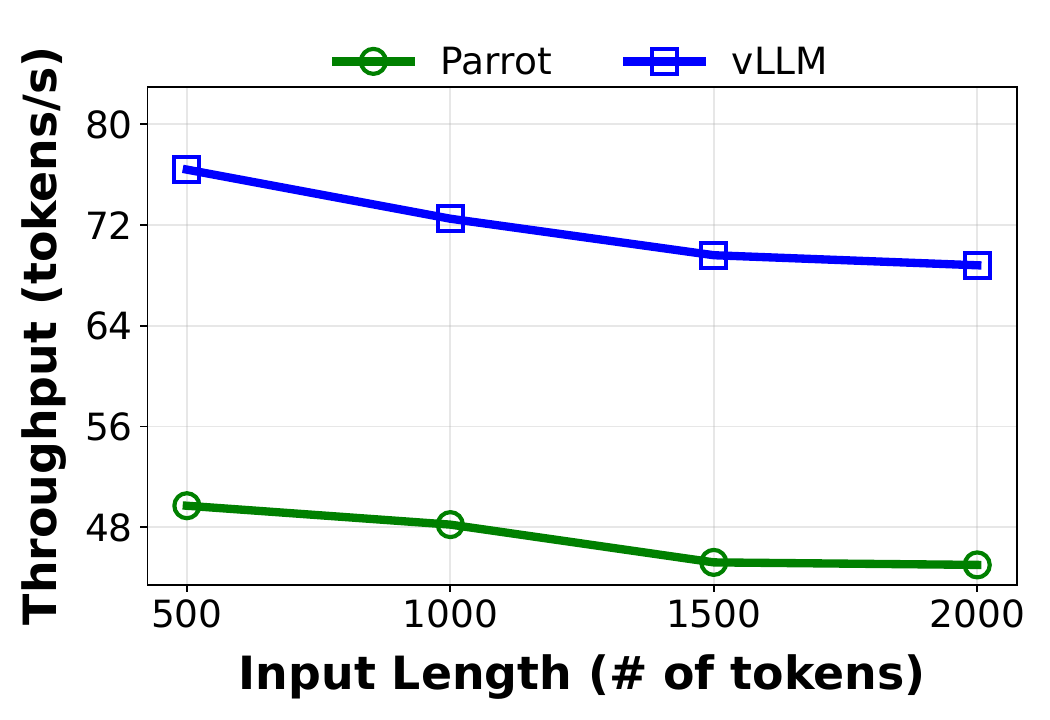}
        % \caption{Throughput vs. Chunk Size}
    \end{subfigure}
    \hfill
    \begin{subfigure}{0.32\columnwidth}
        \includegraphics[width=\linewidth]{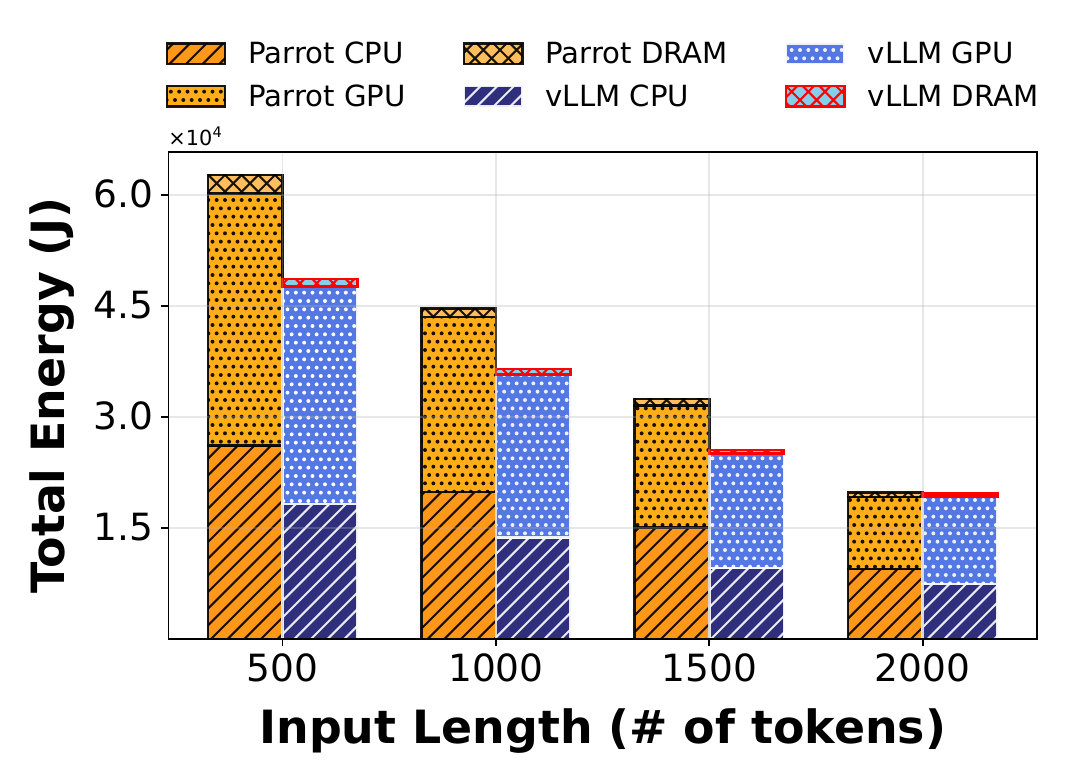}
        % \caption{Energy vs. Chunk Size}
    \end{subfigure}
    
    % Row 2: output length
    \begin{subfigure}{0.32\linewidth}
        \includegraphics[width=\linewidth]{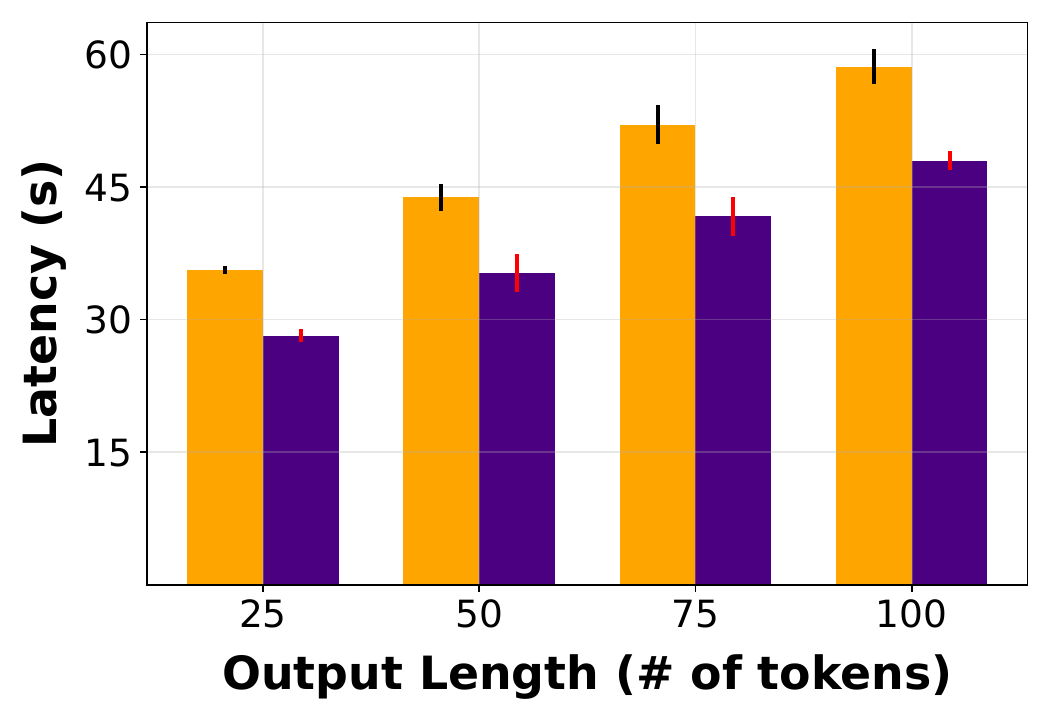}
        % \caption{Latency vs. output length}
    \end{subfigure}
    \hfill
    \begin{subfigure}{0.32\linewidth}
        \includegraphics[width=\linewidth]{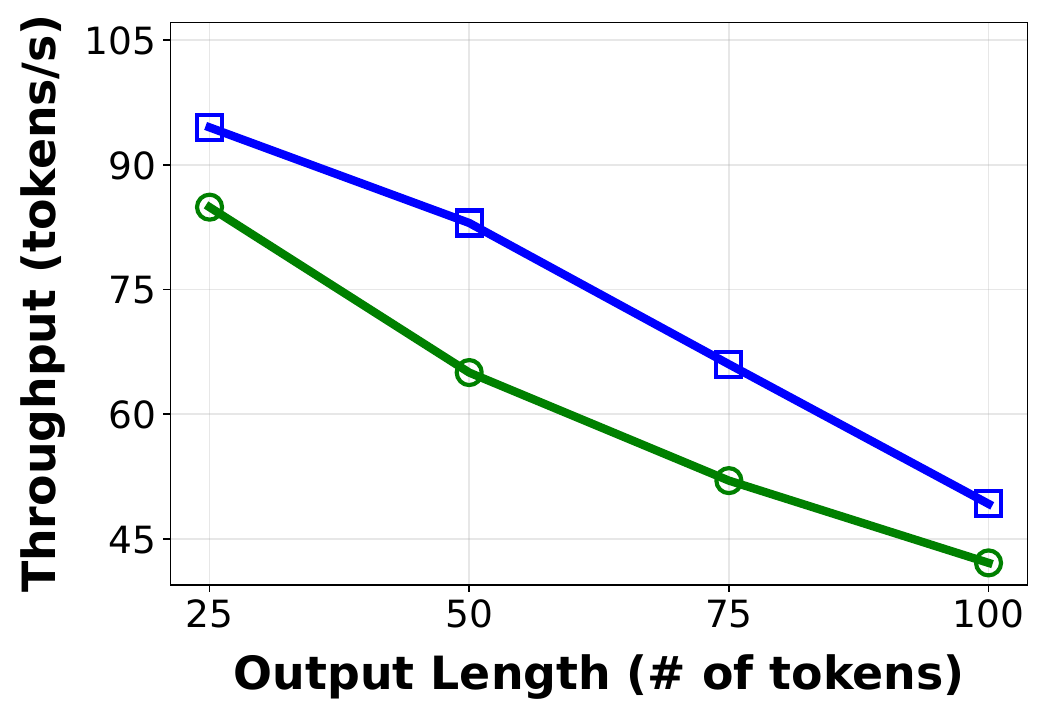}
        % \caption{Throughput vs. output length}
    \end{subfigure}
    \hfill
    \begin{subfigure}{0.32\linewidth}
        \includegraphics[width=\linewidth]{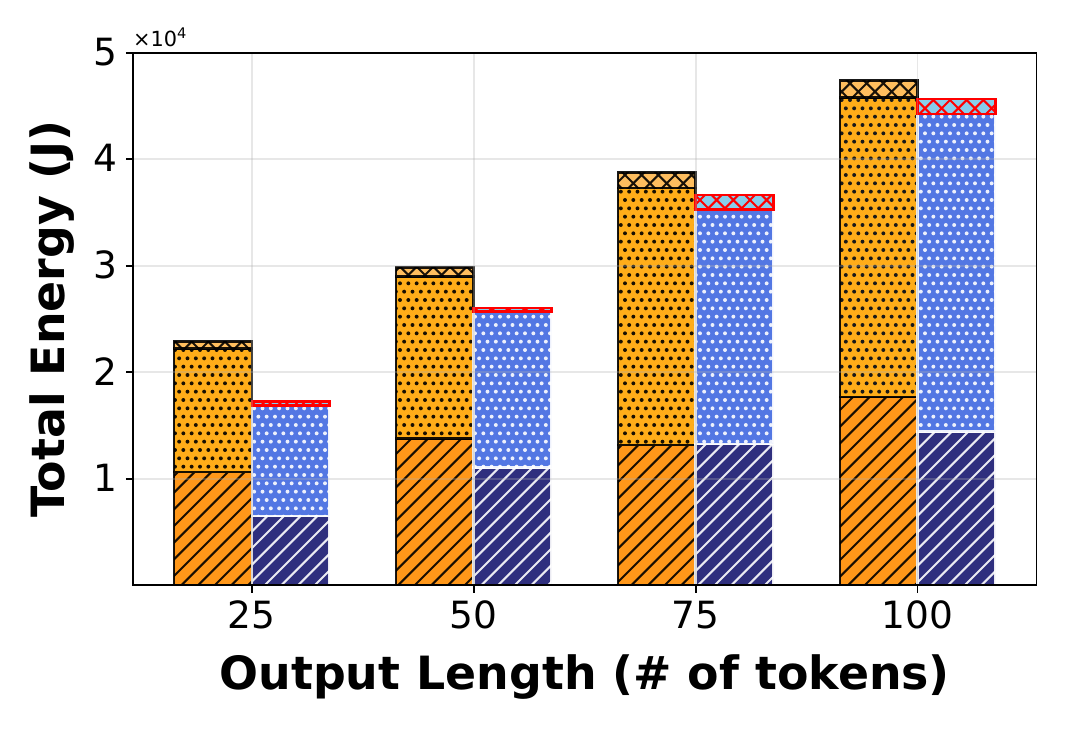}
        % \caption{Energy vs. output length}
    \end{subfigure}
    
    % Row 3: Batch Size
    \begin{subfigure}{0.32\columnwidth}
        \includegraphics[width=\linewidth]{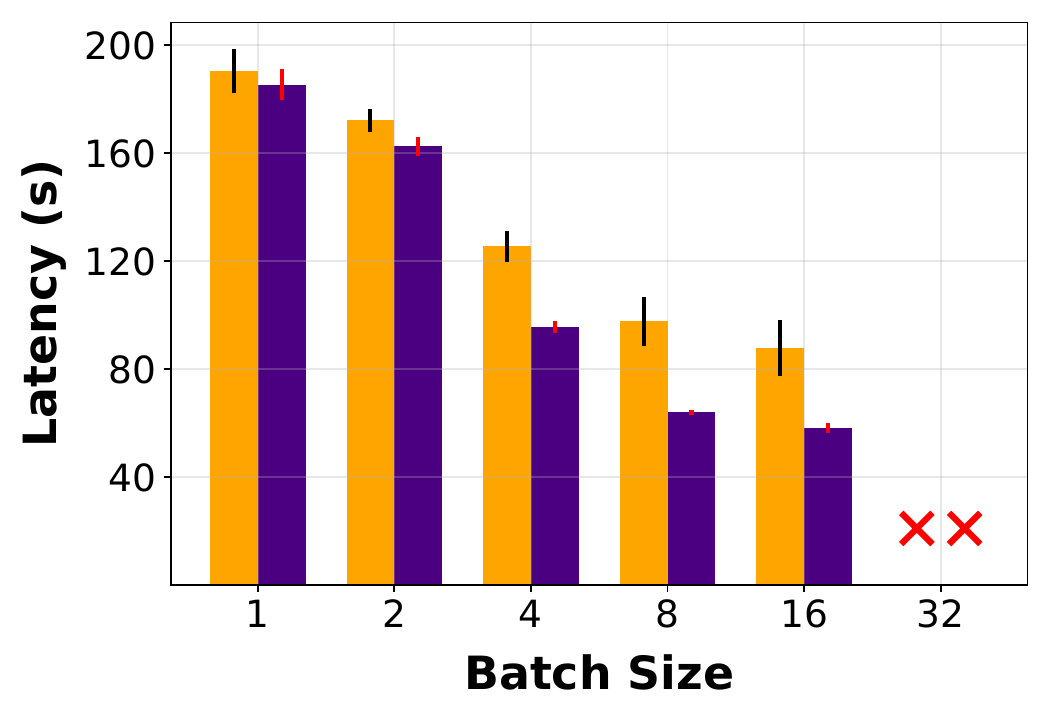}
        % \caption{Latency vs. Batch Size}
    \end{subfigure}
    \hfill
    \begin{subfigure}{0.32\columnwidth}
        \includegraphics[width=\linewidth]{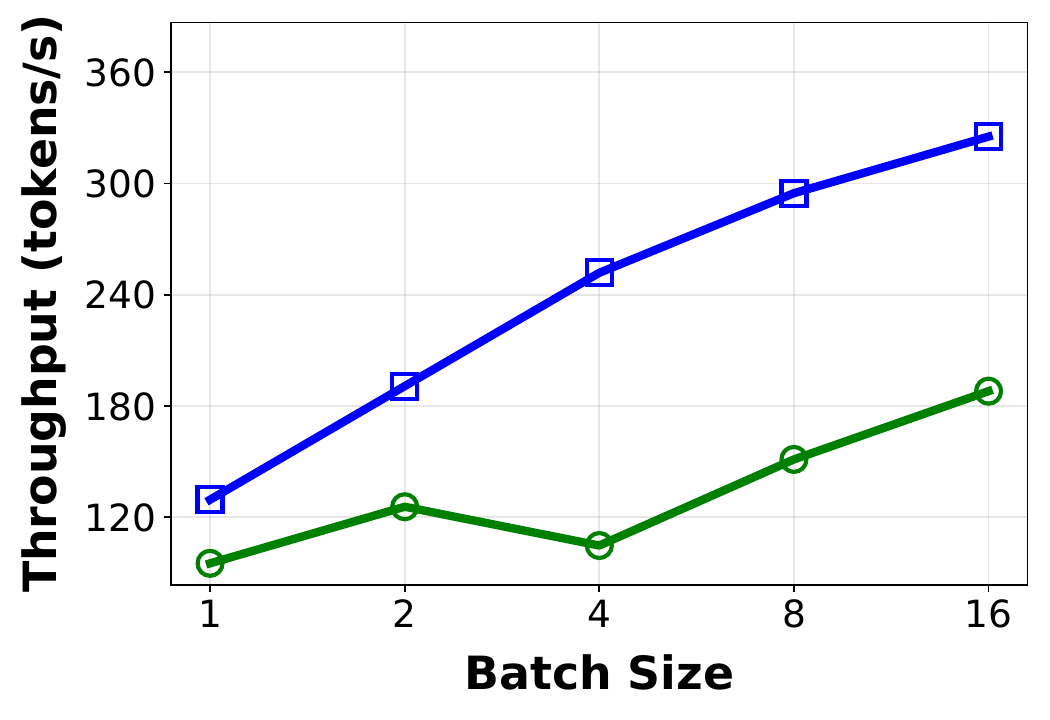}
        % \caption{Throughput vs. Batch Size}
    \end{subfigure}
    \hfill
    \begin{subfigure}{0.32\columnwidth}
        \includegraphics[width=\linewidth]{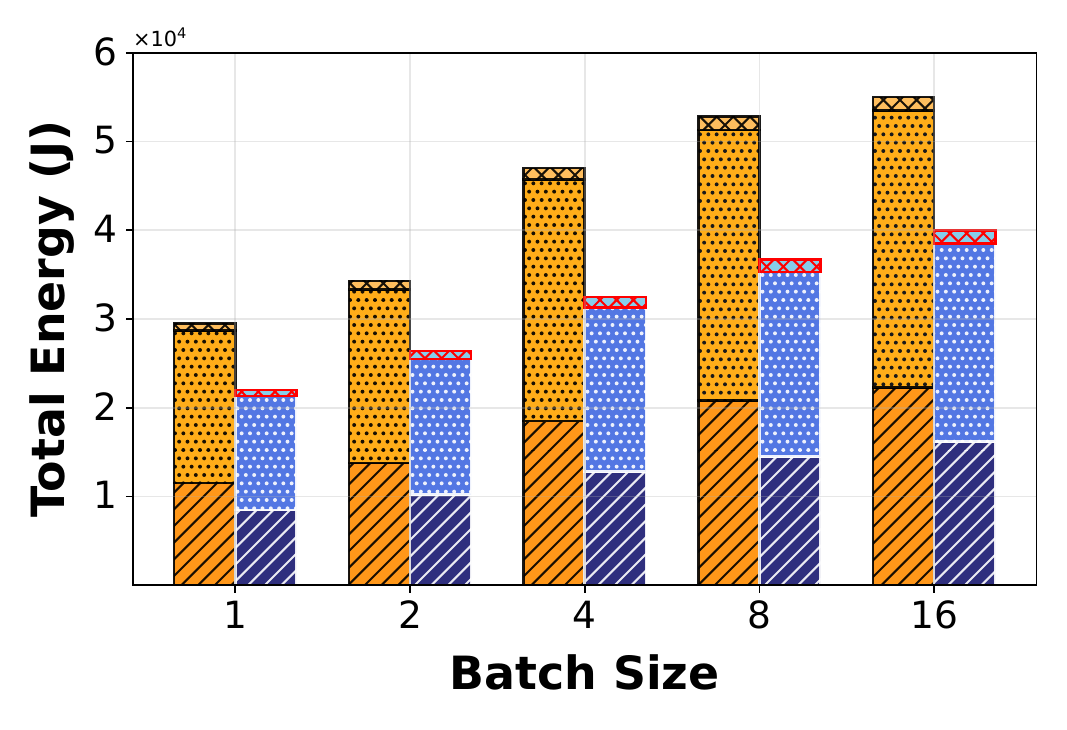}
        % \caption{Energy vs. Batch Size}
    \end{subfigure}
    
    % Row 4: Power Cap
    \begin{subfigure}{0.32\linewidth}
        \includegraphics[width=\linewidth]{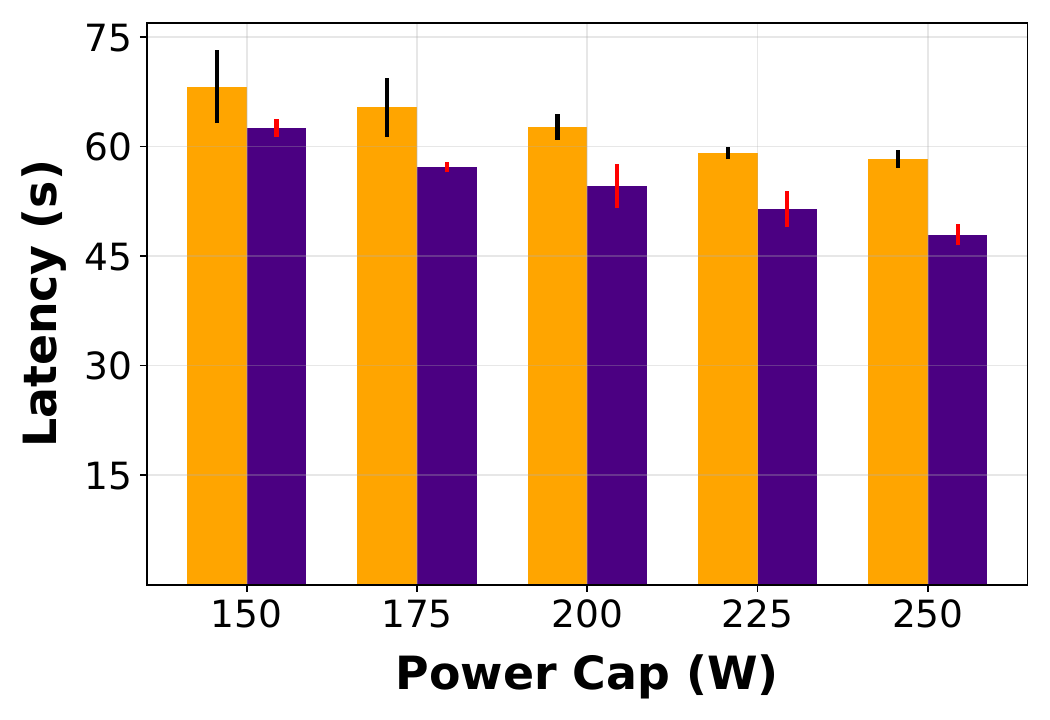}
        % \caption{Latency vs. Power Cap}
    \end{subfigure}
    \hfill
    \begin{subfigure}{0.32\linewidth}
        \includegraphics[width=\linewidth]{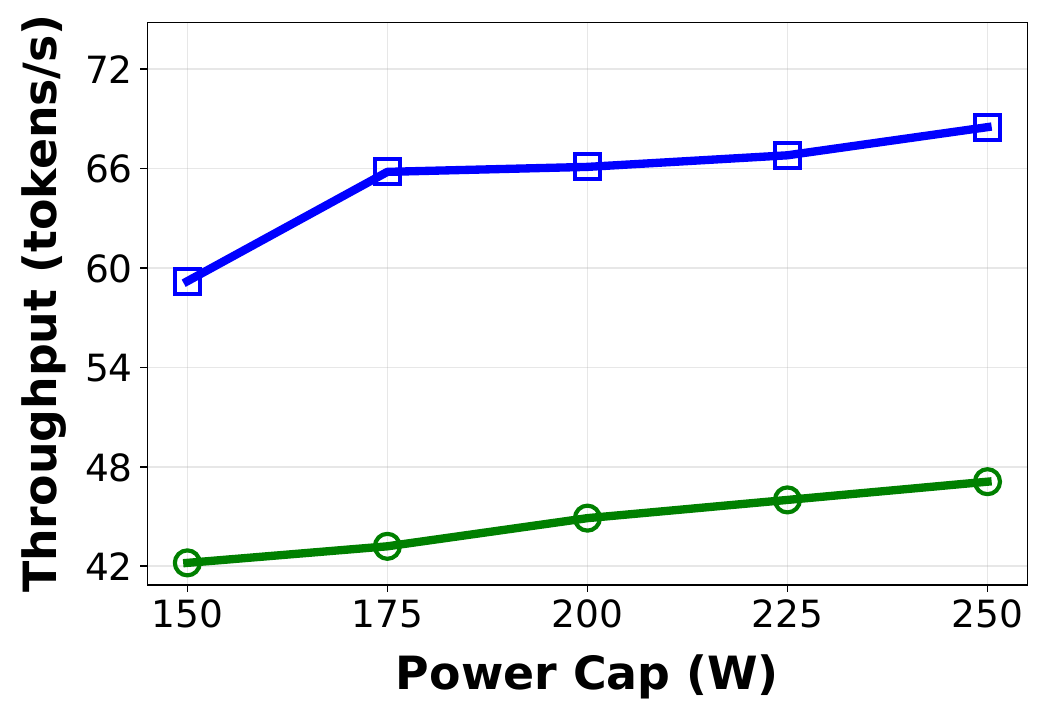}
        % \caption{Throughput vs. Power Cap}
    \end{subfigure}
    \hfill
    \begin{subfigure}{0.32\linewidth}
        \includegraphics[width=\linewidth]{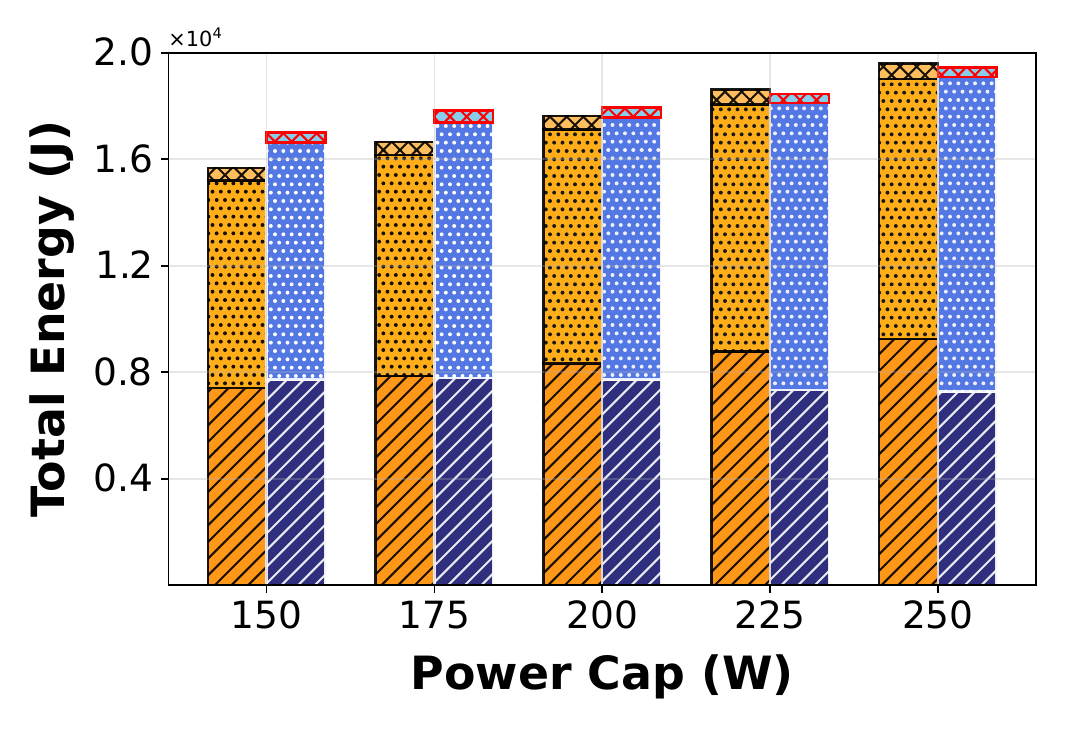}
        % \caption{Energy vs. Power Cap}
    \end{subfigure}
    
    \caption{RQ3 results on chain summarization workload showing impact of input length, output length, batch size, and GPU power capping on latency, throughput, and component-wise (CPU,GPU, and DRAM) energy consumption. The red X markers denote batch sizes that are infeasible due to GPU OOM issue.}
    \label{fig:rq3_chain}
\end{figure}

\noindent\paragraph{Input Length Effect:}
The input length here corresponds to the token count of each document chunk processed during summarization.
% The first row of Figure~\ref{fig:rq3_chain} shows the impact of input length (500--2000 tokens) on vLLM and Parrot across latency, throughput, and energy consumption, where we observed an expected performance trend over the chosen workload. However, the performance across serving systems is interesting. 
In Figure~\ref{fig:rq3_chain}, the first row plots the impact of varying input length (500-2000 tokens). Here, latency decreases as larger input length amortize prefill overhead across fewer LLM calls. Throughput remains relatively stable, since prefill scaling does not directly affect decode throughput. Energy consumption decreases significantly with larger input length, as fewer sequential requests are executed. GPU remains the dominant contributor, with CPU and DRAM following similar trends. vLLM achieves consistently lower energy and latency than Parrot, mainly due to more efficient chunked prefill that overlaps computation with decoding.
% While we observed an expected performance trend over the chosen workload, the performance across serving systems is interesting. 
% vLLM consistently outperforms Parrot across all metrics. Parrot consumes noticeably more energy relative to its achieved throughput due to frequent small prefills and scheduling overheads, which introduce additional computation and consequently higher energy use. vLLM avoids this inefficiency via chunked prefill \cite{agrawal2024taming}, which processes long prompts incrementally while overlapping with decodes, thereby reducing redundant GPU activity.

\noindent\paragraph{Output Length Effect:} 
The second row in Figure~\ref{fig:rq3_chain} shows the impact of varying output length (25-100 tokens). 
Latency and energy rise monotonically as longer generations increase sequential decoding cost, while throughput declines due to extended per-request completion. Unlike the LLM-powered search workload (Figure~\ref{fig:rq1-outsize-tp-eg}), where throughput grows with output length for independent requests, sequential dependencies among LLM calls in chain summarization extend the critical path and reduce overall token throughput. vLLM sustains a clear advantage in shorter outputs, consistent with its kernel-level acceleration (see \ref{comparisonLabel}).

\noindent\paragraph{Batch Size Effect:} In Figure~\ref{fig:rq3_chain}, the third row plots the impact of increasing batch size. We vary \texttt{max-num-seqs} (maximum concurrent requests per iteration) while holding \texttt{max-num-batched-tokens} fixed to isolate request-level batching effects. Larger batches improve throughput and reduce latency until GPU memory limits are reached, but total energy rises due to heavier GPU utilization. vLLM maintains higher throughput and lower latency across all feasible batch sizes through continuous batching. At batch size 16 (the most energy-intensive configuration among all our experiments), vLLM reduces total energy consumption by up to 28\% compared to Parrot. Parrot shows occasional inefficiencies when heterogeneous requests are batched together, reflecting less flexible batching control.

% vLLM sustains higher throughput and lower latency across all batch sizes, benefiting from continuous batching (join in-progress batches), which dynamically admits new requests into in-progress batches and reduces head-of-line blocking. Parrot, in contrast, exhibits an outlier at batch size 4, where throughput gain is negative but GPU energy spikes. This inefficiency arises from misaligned batching of heterogeneous requests, which prevents prefix reuse while incurring queuing overhead. vLLM mitigates this in its recent update (\emph{vLLM V1} \cite{vllm_v1}), which pipelines API processing with asynchronous execution. Smaller batches consume less energy, while larger batches maximize throughput at higher energy cost. At batch size 16--the most energy-intensive configuration among all our experiments--vLLM reduced total energy consumption by up to 28\% compared to Parrot. 

\noindent\paragraph{GPU Power Capping Effect:}
The fourth row in Figure~\ref{fig:rq3_chain} plots the effect of GPU power capping (150W to 250W). Lower power caps moderately reduce total energy but increase latency and reduces throughput. Parrot achieves larger relative energy savings (up to 21\% when decreasing the power cap from 250W to 150W) compared to vLLM (12.6\%), as its workflow-level scheduling distributes computation more evenly across CPU and GPU under constrained power. However, vLLM sustains higher throughput and lower latency across all caps, reflecting better GPU efficiency under kernel-level optimizations. 

% FlashAttention and chunked prefill allow vLLM to maintain smoother utilization even at reduced power budgets, whereas Parrot experiences sharper throughput degradation. Reducing GPU power limit curtails total energy draw but comes at the expense of increased latency and reduced throughput. 

% The magnitude of this trade-off aligns with prior findings dealing with LLM inference requests separately, where moderate caps offer meaningful energy savings with tolerable performance penalties, but aggressive capping (e.g., 150W) amplifies latency overheads disproportionately \cite{samsi2023words, stojkovic2025dynamollm}. \textcolor{red}{this is good to bring existing reports on single-request tasks but again for all four knobs, what's similar and what's new? we may point that in the final insight bix. -- Importantly, these trends mirror what has been reported for single-request inference but the impacts are more pronounced in multi-request workflows. This amplification occurs because sequential dependencies compound the slowdown across multiple calls, stretching end-to-end latency and diminishing throughput more severely than in isolated single requests.}

% \begin{figure}[!hbtp]
%     \centering
%     \includegraphics[width=0.6\linewidth,height=0.4\textheight,keepaspectratio]{pictures/compare_util.pdf}
%     \caption{GPU utilization comparison between vLLM and Parrot for chain summarization workload (batch size 16). Values are averaged across 10 runs.}
%     \label{fig:gpu-utilization}
% \end{figure}

\noindent\paragraph{Serving System Comparison:}\labeltext{\texttt{Serving System Comparison}}{comparisonLabel}
The two serving systems embody contrasting optimization philosophies that shape the trends in Figure \ref{fig:rq3_chain}. vLLM prioritizes \textit{engine-level efficiency}, integrating PagedAttention \cite{kwon2023efficient}, FlashAttention \cite{dao2022flashattention}, chunked prefill \cite{agrawal2024taming} and continuous batching \cite{yu2022orca} to sustain high GPU utilization and minimize kernel-launch overheads. PagedAttention reduces KV-cache fragmentation by managing memory as paged blocks, while FlashAttention lowers memory traffic during attention computation. Chunked prefill further improves efficiency for long prompts by splitting them into smaller segments that can be processed incrementally and overlapped with decoding, amortizing prefill cost and reducing GPU stall time. Continuous batching complements this by dynamically admitting new requests into active batches, reducing queuing delays.
\begin{wrapfigure}{r}{0.45\linewidth}
  \centering
  % \vspace{-10pt} % adjust vertical placement if needed
  \includegraphics[width=0.9\linewidth,height=0.8\textheight,keepaspectratio]{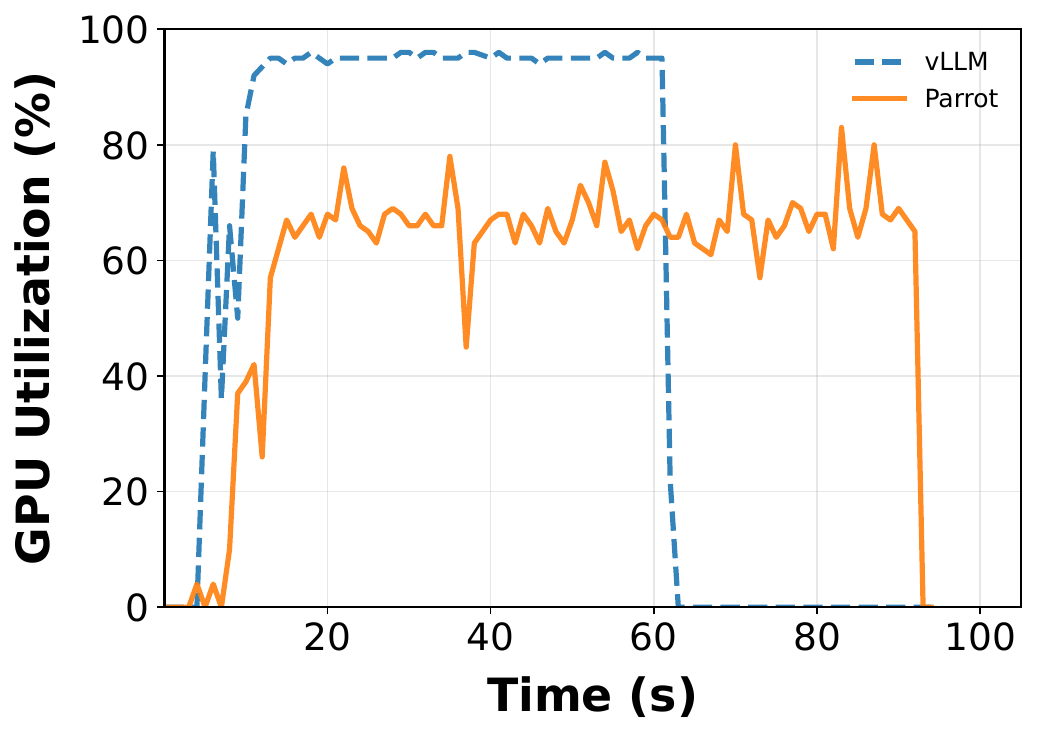}
  \caption{GPU utilization comparison between vLLM and Parrot for chain summarization workload (batch size 16). Values are averaged across ten runs.}
    \label{fig:gpu-utilization}
  \vspace{-20pt}
\end{wrapfigure}
Together, these optimizations collectively enable vLLM to maintain higher and more stable utilization (mean 90.1\%, peak 96.0\%) compared to Parrot (mean 63.3\%, peak 83.0\%), as shown in Figure~\ref{fig:gpu-utilization}. Although vLLM sustains higher GPU utilization, its total energy remains lower because these cycles are efficiently used by reducing idle stalls and shortening latency. The shorter execution time offsets higher instantaneous power, so the overall energy (power $\times$ time) remains lower despite higher utilization. 

% Together, these optimizations collectively enable vLLM to maintain higher and more stable utilization (mean 90.1\%, peak 96.0\%) compared to Parrot (mean 63.3\%, peak 83.0\%), as shown in Figure~\ref{fig:gpu-utilization}, resulting in lower latency and total energy despite higher instantaneous power. The shorter runtime offsets the increased power draw, resulting in overall energy savings.

Parrot, in contrast, adopts a \textit{workflow-aware optimization} paradigm through semantic variables \cite{lin2024parrot}, which expose dependency graphs among LLM calls for DAG-level co-scheduling. This approach improves coordination across dependent stages and provides relative gains under aggressive power caps by balancing computation across CPU and GPU resources. However, Parrot does not explicitly mitigate KV-cache fragmentation or optimize fine-grained kernel execution, leading to reduced GPU occupancy at larger batch sizes and higher cumulative energy in sequential workloads (as shown in Figure \ref{fig:rq3_chain}).

Overall, the comparison highlights a broader trade-off between hardware-centric throughput optimization (vLLM) and application-centric workflow scheduling (Parrot). While vLLM’s kernel- and memory-level optimizations provide superior energy efficiency under decode-heavy conditions, Parrot’s DAG-level scheduling achieves modest gains under constrained power budgets, consistent with prior analyses of inference-time efficiency optimizations \cite{fernandez-etal-2025-energy}.

\findingsbox{Our comparison between state-of-the-art serving systems vLLM (optimized for engine-level efficiency) and Parrot (optimized for workflow-aware scheduling) identifies a fundamental design tension in multi-request serving. vLLM’s low-level optimizations, including PagedAttention, FlashAttention, and chunked prefill, sustain consistently higher GPU utilization and reduce stall time, resulting in up to 28\% lower total energy under decode-heavy, sequential workloads. In contrast, Parrot’s semantic-variable scheduling offers modest benefits only under aggressive power caps, trading GPU saturation for CPU-GPU balance. This result highlights that fine-grained engine-level control over GPU execution often outweighs high-level workflow semantics in determining efficiency. Looking ahead, future serving designs should either (i) prioritize GPU stall elimination to maximize efficiency in decode-bound workflows \cite{lin2025bullet}, or (ii) integrate both paradigms in a dynamic, workload-aware manner to combine the complementary strengths of engine-level and workflow-level optimization \cite{su2025efficient}.}

\section{Discussion}
\label{sec:discussion}

% \noindent{\textbf{Limitations of Our Study:}}
This work establishes a baseline characterization of performance--energy trade-offs in multi-request LLM inference, providing empirical foundations for sustainable system design. Although comprehensive, we acknowledge the following limitations in our study. 

% First, we restrict our exploration of performance--energy trade-offs to three representative energy related knobs to isolate their individual effects. Other influential knobs, such as model-level parameters (e.g., quantization, parallelism, activation sparsity) and system-level factors (e.g., multi-GPU scaling, CPU-GPU co-scheduling), are excluded to maintain controlled and reproducible single-GPU evaluations. These excluded knobs interact in complex, often non-linear ways with workload parameters, producing cascading impacts on efficiency \cite{rajput2025tu}. While we justify our focus on the three most critical and widely exposed controls in Section \ref{sec:background}, future work will extend this characterization to joint tuning of model-, system-, and workload-level knobs under distributed and heterogeneous settings. 

First, we restrict our exploration of performance--energy trade-offs to three representative energy-related knobs to isolate their individual effects. Other influential knobs, such as model-level parameters (e.g., quantization, parallelism, activation sparsity) and system-level factors (\eg multi-GPU scaling, CPU--GPU co-scheduling), are excluded to maintain controlled and reproducible single-GPU evaluations. These excluded knobs interact in complex, often non-linear ways with workload parameters, producing cascading impacts on efficiency \cite{rajput2025tu}. \Update{In particular, quantization techniques (e.g., FP16, INT8, or lower-precision formats) are widely used in practice to reduce memory bandwidth pressure and per-token energy consumption, especially during the decode phase. However, quantization may also introduce secondary effects such as increased CPU-side overhead, altered batching behavior, reduced KV-cache efficiency, and, in some cases, longer per-token decode latency depending on kernel maturity and hardware support. While we justify our focus on the three most critical and widely exposed controls in Section~\ref{sec:background}, future work will extend this characterization to joint tuning of model-, system-, and workload-level knobs, including quantization-aware inference, under distributed and heterogeneous settings.}

Second, we restrict our study to Llama 2-7B because it represents a widely used open-source, decoder-only baseline with transparent licensing and stable inference behavior, which is an essential requirement for controlled, reproducible energy measurements. Parrot (at the time of experimentation) does not yet support grouped-query attention (GQA) \cite{ainslie2023gqa} or mixture-of-experts (MoE) \cite{shazeer2017sparsely} routing, which are core to newer families such as Llama 3, Qwen 3, and DeepSeek R1. Therefore, evaluating such modern architectures along with large reasoning models (LRMs) remain an important future direction. We plan to incorporate complementary serving systems (e.g., Hugging Face TGI, SGLang) that natively support these architectures to extend our comparative analysis beyond vLLM and Parrot. 
% Second, Parrot currently lacks support for grouped-query attention (GQA) \cite{ainslie2023gqa}, preventing us from evaluating more recent and widely adopted open-source models such as Llama 3 and Mistral, which employ GQA for efficiency. This limits the generalizability of our findings to newer model generations. \textcolor{purple}{In future work, we plan to incorporate complementary serving systems that natively support GQA (e.g., Hugging Face TGI, SGLang) to extend our comparative analysis beyond vLLM and Parrot.}

% Third, our evaluation is conducted on a single NVIDIA A100 GPU to ensure controlled, repeatable, and fine-grained energy characterization. This setup enables precise attribution of component-level energy without interference from distributed communication overheads. While this single-node design isolates system-level effects, we acknowledge that production LLM inference typically operates on multi-GPU, multi-node clusters. Extending this study to distributed environments and newer accelerators (e.g., NVIDIA H100, RTX A6000) will be essential to capture interconnect energy costs and scaling dynamics under realistic deployment settings.

Finally, our evaluation is conducted on a single NVIDIA A100 GPU to ensure controlled, repeatable, and fine-grained energy characterization. This setup enables precise attribution of component-level energy without interference from distributed communication overheads. While this single-node design isolates system-level effects, we acknowledge that production LLM inference typically operates on multi-GPU and multi-node clusters. \Update{In such environments, interconnect communication over NVLink, PCIe, or InfiniBand, as well as cross-device synchronization, model-parallel coordination, and distributed scheduling, can contribute non-trivial energy overheads that are not captured in our measurements. Extending this study to distributed environments and newer accelerators (\eg NVIDIA H100, RTX A6000) would enable characterization of these additional energy dynamics and reveal how communication and computation jointly shape end-to-end efficiency under realistic deployment settings.}

\section{Conclusion}
\label{sec:conclusion}

This paper presents a comprehensive assessment of performance--energy trade-offs in multi-request LLM workflows. By deploying four representative workloads and systematically varying key energy knobs, we expose how workload- and system-level decisions shape both efficiency and responsiveness. For the chosen serving systems, vLLM and Parrot, we find that engine-level optimizations like continuous batching and PagedAttention offer significant gains in GPU utilization and energy efficiency, while workflow-aware scheduling provides relative benefits under constrained power budgets. Some of the key insights include, batch size is the most influential knob, but its effects are highly workload-dependent. The output length induces predictable, near-linear energy scaling, offering limited efficiency improvements. GPU power capping can provide modest, controllable savings, but aggressive limits amplify latency penalties. Looking ahead, dynamically adjusting GPU power caps through fine-grained 
% dynamic voltage and frequency scaling (DVFS) 
DVFS and automating the tuning of model-, workload-, and system-level energy knobs using reinforcement learning-based approaches represent promising directions toward adaptive, energy-aware LLM serving.

\bibliographystyle{plain}
\bibliography{refs}

% \received{October 2025}
% \received[revised]{December 2025}
% \received[accepted]{January 2026}

\begin{comment}
\section{Appendices}

If your work needs an appendix, add it before the
``\verb|\end{document}|'' command at the conclusion of your source
document.

%%
%% If your work has an appendix, this is the place to put it.
\appendix

\section{Research Methods}

\subsection{Part One}

Lorem ipsum dolor sit amet, consectetur adipiscing elit. Morbi
malesuada, quam in pulvinar varius, metus nunc fermentum urna, id
sollicitudin purus odio sit amet enim. Aliquam ullamcorper eu ipsum
vel mollis. Curabitur quis dictum nisl. Phasellus vel semper risus, et
lacinia dolor. Integer ultricies commodo sem nec semper.

\end{comment}

\end{document}